\documentclass[conference,letterpaper]{IEEEtran}

\usepackage{graphicx,color}
\usepackage{algorithmic}
\usepackage{amsmath,bbm,amssymb,amsfonts,amstext,amsopn,sansmath}

\newcommand{\RED}{\color[rgb]{0,0,0}}
\newcommand{\BLUE}{\color[rgb]{0,0,0}}
\newcommand{\GREEN}{\color[rgb]{0,0,0}}
\usepackage[utf8]{inputenc} 
\usepackage[T1]{fontenc}
\usepackage{url}
\usepackage{ifthen}
\usepackage{cite}
\usepackage{amsmath} 

\usepackage{psfrag}	
\usepackage{pstool}
\usepackage{etoolbox}

\newcommand{\Nul}{\mathcal{N}_{\mathrm{ul}}}
\newcommand{\Nlb}{\mathcal{N}_{\mathrm{lb}}}

\newcommand{\Nset}{\mathcal{N}}
\newcommand{\Lset}{\mathcal{L}}

\newcommand{\Llb}{\boldsymbol{\ell}_{\Nlb}}

\newcommand{\Edg}{\mathbf{\bf e}}
\newcommand{\Lbl}{\boldsymbol{\ell}}

\newcommand{\Loss}{\mathcal{C}}

\newcommand{\E}{\mathbb{E}}

\newcommand{\ins}{\mathrm{in}}
\newcommand{\out}{\mathrm{out}}

\newtheorem{theo}{Theorem}

\newtheorem{remk}{Remark}
\newtheorem{defin}{Definition}

\newtheorem{corol}{Corollary}

\begin{document}
\title{R\'{e}nyi Entropy Bounds on the Active Learning Cost-Performance Tradeoff} 


\author{%
  \IEEEauthorblockN{Vahid Jamali}
  \IEEEauthorblockA{Dept. of Electrical Engineering\\
                    University of Erlangen-Nuremberg, Germany\\
                    vahid.jamali@fau.de \vspace{-0.7cm}}
  \and
  \IEEEauthorblockN{Antonia Tulino}
  \IEEEauthorblockA{Dept. of Electrical Engineering\\
                    University of Napoli Federico II, Italy\\
                    antoniamaria.tulino@unina.it \vspace{-0.7cm}}
  \and
  \IEEEauthorblockN{Jaime Llorca and Elza Erkip}
  \IEEEauthorblockA{Tandon School of Engineering\\
                    New York University, New York\\
                     \{jllorca, elza\}@nyu.edu \vspace{-0.7cm}}
                 
}


\maketitle

\begin{abstract}
Semi-supervised classification, one of the most prominent fields in machine learning, studies how to combine the statistical knowledge of the often abundant unlabeled data with the often limited labeled data in order to maximize overall classification accuracy. 
In this context, the process of actively choosing the data to be labeled is referred to as {\em active learning}. 
In this paper, we initiate the non-asymptotic analysis of the optimal policy for semi-supervised 
classification with {\em actively obtained labeled data}. 
Considering a 
general Bayesian classification model, 
we provide the first characterization of the jointly optimal active learning and semi-supervised classification policy,  
in terms of the cost-performance tradeoff driven by the label query budget (number of data items to be labeled) and overall classification accuracy. 
Leveraging recent results on the R\'{e}nyi  Entropy,  
we  derive tight information-theoretic bounds on such active learning cost-performance tradeoff. 
\end{abstract}


\section{Introduction}

In many 
classification problems, the cost of obtaining labeled data can be very high, for example when expert knowledge and/or human intervention is required (e.g., labeling objects in images or videos, obtaining personal data, or performing medical tests). 
In this highly common setting, the resulting classification problem pertains to 
the field of {\em semi-supervised learning}, which studies how to combine the distribution of the often abundant unlabeled data with the often limited labeled data in order to aid the classification process \cite{chapelle2006semi,zhu2005semi}. 
A critically related problem is that of 
{\em active learning}, which focuses on 
querying the labels of 
the limited-size set of 
data items 
that 
most significantly improves overall classification accuracy \cite{settles2012active,Aggarwal_2014_chapter22}.
The basic premise of active learning (AL) is that if the cost of obtaining labels is high, we can hope to achieve our learning objective with less overall cost by taking control of the labeling process. 

Existing works in the AL literature can be distinguished along two main dimensions: (i) the heuristic vs. theoretic nature of the associated label query strategy, and (ii) the nature of the input/observed data. 
Regarding (i),  
heuristic based approaches use a variety of intuitive criteria to label the items that are most informative about the decision boundary of the learned model. Examples include  heterogeneity-based models 
performance-based models, 
and representativeness-based models. 
An excellent survey on active learning and its various heuristic techniques, applications, and model extensions can be found in \cite{settles2012active,Aggarwal_2014_chapter22,elahi2016survey,tuia2011survey,Moore2011community,mirabelli2018active}. 
While such heuristic methods yield flexible algorithms that have shown to provide accurate classifiers with less
cost than that of (passive) supervised methods in specific settings, they cannot provide generalizable performance guarantees. 
On the theoretical front, a few recent works have studied the AL problem assuming an underlying statistical model and providing performance guarantees. However, such works focus on specific models, typically graph models such as the stochastic block model (SBM), and their performance analysis is  only valid in the asymptotic regime where the number of data items goes to infinity \cite{zhao2017survey,abbe2016exact,mossel2015consistency,zhang2014phase,gadde2016active}. 
As a result, while one would expect a significant performance boost when exploiting the knowledge of the underlying statistical model, their results tend to be pessimistic and with limited  practical insight. 
Regarding (ii), 
existing AL methods typically focus on a given family of classification problems depending on the nature of the observed data: from directly observing individual items' features or functions of individual items' features, such as in image classification/segmentation problems~\cite{Zhang2016}, to observing only pairwise interactions (e.g., similarities between pairs of items' features and/or labels) such as in graph-based community detection~\cite{guillory2009label,gu2012towards,zhu2003combining,cesa2013active,dasarathy2015s2}, 
or even observing tuple interactions between multiple items' features and/or labels, as in hypergraph clustering~\cite{zhou2007hypergraphs}.

In this work, we address the design and analysis 
of {\em optimal  AL algorithms for general Bayesian classification problems 
in practical non-asymptotic regimes of the system parameters}. 
Rather than the scaling of the  number of queries 
required to achieve perfect classification, we are interested in the cost-performance tradeoff dictated by the available query budget and overall classification accuracy.
Starting from a general Bayesian classification framework 
that includes as special cases the {SBM} and its variants,
we provide
(i) the first formal characterization of the jointly optimal active learning and semi-supervised classification policy,
as well as (ii) tight information-theoretic bounds on the associated cost-performance tradeoff 
leveraging recent results on R\'{e}nyi  Entropy \cite{Sason_Verdu_MHyp_2018}.

\section{Bayesian Classification Setup}
\label{Sec:Generalmodel}

In this section, we introduce a generic Bayesian classification model that will be the basis for the study of active learning in a variety of settings including those considered in existing AL literature. 

Consider a collection of $N=|\mathcal N|$ data items. 
Each item $n\in\mathcal N$ is associated with a random pair $(\mathbf X_n,L_n)$, 
where $\mathbf X_n\in \mathbb R^d$ and $L_n\in\mathcal L$ denote the feature vector and the label of item $n$, respectively. 
Let $\mathbf{x}_{\mathcal N}\in\mathbb R^{d\times N}$ denote a realization of the collection of data items' features and ${\Lbl}_{\mathcal N}\in\mathcal L^N$  a realization of the associated labels. 
We assume that the relationship between the overall set of features and labels is described by an unknown probability distribution $f(\bf{x}_{\mathcal N},{\Lbl}_{\mathcal N})$.  
In an unsupervised setting, the goal is to estimate the data labels ${\Lbl}_{\cal N}$ from the observation of a (possibly random) function of the features and labels 
$\mathbf{e} \triangleq g(\bf{x}_{\mathcal N},{\Lbl}_{\mathcal N})$, 
where the relationship between the observed function of features and labels and the actual labels is described by a known probability distribution $f(\mathbf{e}, {\Lbl}_{\mathcal N})$.  
We refer to the random variables $\mathbf{E}$ as the {\em observables} and to their realizations $\mathbf{e}$ as  the {\em observations}.
In an AL setting, the observations can be augmented with a properly chosen subset of the items' labels.\footnote{For ease of exposition, we use ${\Lbl}$ and ${\Lbl}_{\mathcal N}$ interchangeably to refer to the entire label vector, and specify $\boldsymbol{\ell}_{\mathcal A}$ when referring to the labels of a subset of items $\mathcal A\subset\mathcal N$. In addition, we assume that $\boldsymbol{\ell}_{\mathcal A}$ contains both the identities and associated labels of the items $\mathcal A\subset\mathcal N$.}

\begin{figure}
		\centering 
		\includegraphics[width=0.45\textwidth]{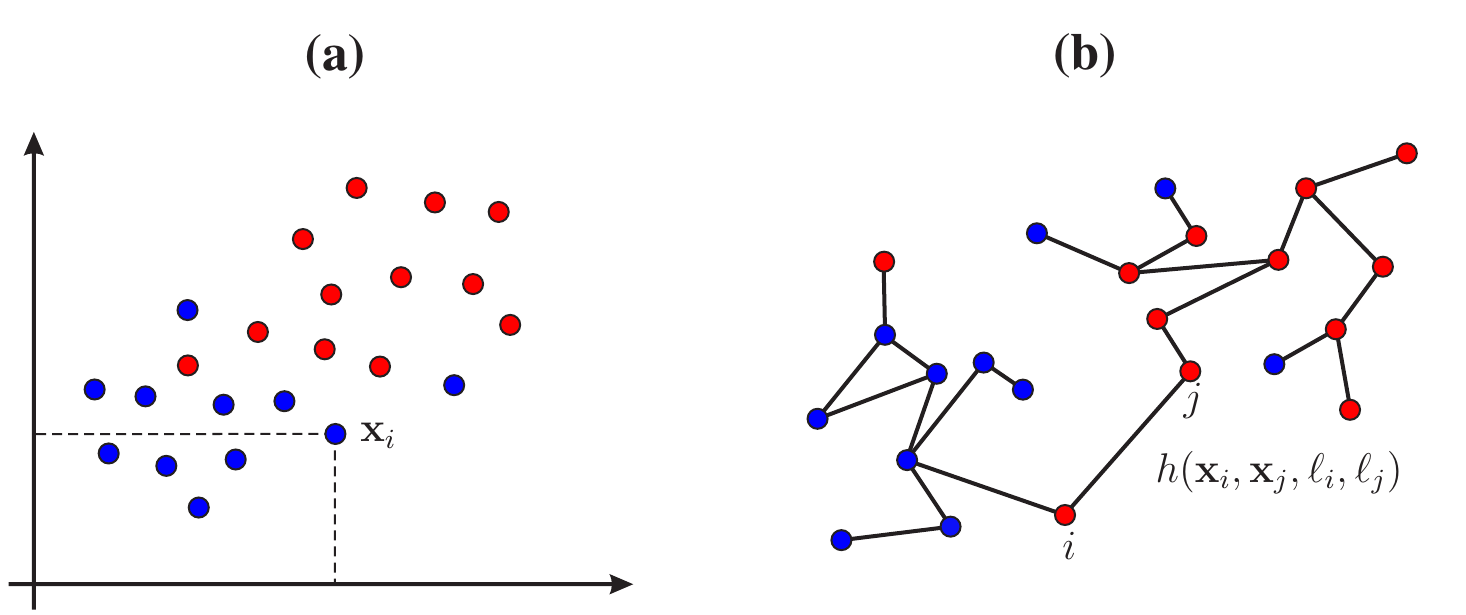}
		\caption{(a) Observable feature space. (b) Observable graph.}
		\vspace{-0.3cm}
		\label{feature_graph} 
\end{figure}

The aforementioned model is simple and comprehensive. It encompasses, as we argue below, a vast class of relevant models that can be classified in terms of the nature of the observed data: from directly observing individual items' features or functions of individual items' features { (as illustrated in Fig. \ref{feature_graph}a}, 
to observing only pairwise interactions 
(as illustrated in Fig. \ref{feature_graph}b, 
or even observing tuple interactions between multiple items' features and/or labels, as in hypergraph clustering. 
Immediate examples of \textit{pairwise interactions (random graphs)} are scenarios where,  
while we may not be able to directly observe the data items' features, we may have access to pairwise functions of two data items' features and labels (e.g., similarities between pairs of data items). 
Such setting is commonly modeled via a random graph, whose nodes represent the collection of data items $\mathcal N$, each associated with a random pair $(\mathbf X_n, L_n)$, and whose random edges represent the observed pairwise functions of the items' feature-label pairs. In particular, the observable $\mathbf{E}=g(\bf{X}_{\mathcal N},{\bf L}_{\mathcal N})$ can be described by an $N \times N$ binary random matrix 
 where element $(i,j)$ is given by $\mathbf{E}_{i,j} = h(\mathbf{X}_i,\mathbf{X}_j, L_i, L_j )$, with $h(\cdot)$ being a properly defined (possibly noisy) function. 
 A renowned example is given by the SBM \cite{zhao2017survey,abbe2016exact,mossel2015consistency,zhang2014phase,gadde2016active}, a popular random graph model for community detection that generalizes the well known Erd{\"o}s-Renyi model.
In this case, $h(\mathbf X_i,\mathbf X_j, L_i, L_j)$ is a noisy binary function defined as  
\begin{eqnarray}
h(\mathbf X_i,\mathbf X_j, L_i, L_j)=    \left \{
\begin{array}{ccc}
1 &\qquad &  {\rm if} \,\,\,   U_{i,j} \leq \delta_{i,j}\\
0 &\qquad &   { \rm otherwise} , 
\end{array}
\right., 
\label{functionlink0}
\end{eqnarray}
with 
$\delta_{i,j} =q_{\ins}{\bf 1}\{  L_i=L_j  \} +q_{\out}{\bf 1}\{  L_i\neq L_j  \}$,
with $U_{i,j} \sim \mathcal{U}(0,1)$ independent across $(i,j)$, and  $0<q_{\out}<q_{\ins} <1 $.
Similar examples can be provided for the
the geometric block model (GBM) \cite{galhotra2019GBM},  the Gaussian mixture block model (GMBM) \cite{abbe2018GMBM}, and the Euclidean random graph (ERG) \cite{sankararaman2018ERG}.


\section{Optimal Semi-supervised Classifier}

Starting from the general model of Sec. \ref{Sec:Generalmodel}, 
let 
$	\Loss :  (\hat{\boldsymbol{\ell}},{\boldsymbol{\ell}}) \in \mathcal L^2 \rightarrow   \Loss(\hat{\boldsymbol{\ell}}, {\boldsymbol{\ell}}) \in \mathbb R$ 
denote the loss function that quantifies the cost incurred when the true realization of $\mathbf{L}$ is $\boldsymbol{\ell}$ while the 
chosen estimate is $\hat{\boldsymbol{\ell}}$. 
In an unsupervised setting, the Bayesian estimator is the one that minimizes the  conditional  risk 
$\mathcal{R}\big (\hat{\boldsymbol{\ell}}|  \mathbf{e}\big) \triangleq \E\big[\Loss(\hat{\boldsymbol{\ell}}, {\mathbf{L}})| \mathbf{E} = \mathbf{e}\big]$, where the expectation is with respect to the conditional probability $f_{\mathbf{L}|\mathbf{E}}(\boldsymbol{\ell}
|\mathbf{e})$.
In the context of active learning, the  observations  $\mathbf{e}$, 
which can be thought of as passively-obtained  data,  are augmented with a set of actively-obtained data $\mathbf{s}$, 
referred to as the {\em side information}.  
Letting $\mathbf{y} =(\mathbf{e},\mathbf{s})$ denote the augmented observations,  
the Bayesian estimator now minimizes the conditional risk $
\mathcal{R}\big (\hat{\boldsymbol{\ell}}|  \mathbf{y}\big) =
\E\left[\Loss(\hat{\boldsymbol{\ell}}, {\mathbf{L}})| \mathbf{Y} = \mathbf{y}\right]$.
If the side information $\mathbf{s}$ is a 
subset of the labels $\boldsymbol{\ell}$,  minimizing the conditional risk corresponds
to solving  a \textit{semi-supervised classification} problem and  the procedure of choosing a subset of the elements of $\boldsymbol{\ell}$ as the side information $\mathbf{s}$ is referred to as \textit{active learning}.


Formally, let $\Nlb$ denote the set of labeled items, and $\Llb$ their associated labels. 
Similarly, let $\Nul \triangleq \Nset\backslash \Nlb$ denote the set of unlabeled items we wish to classify, and $\hat{\Lbl}_{\Nul}$ their label estimates. 
The optimal semi-supervised classifier is given by
\begin{eqnarray}\label{Eq:OptClassification} 
	\hat{\Lbl}_{\Nul}^{*} = \underset{\hat{\Lbl}_{\Nul}\in \Lset^{|\Nul|} }{\mathrm{argmin}}\,
	 \mathcal{R}\big (\hat{\Lbl}_{\Nul}|   \Llb, \Edg  \big).
\end{eqnarray}

In the next subsection, we focus on the design of
the optimal active learning policy for the above semi-supervised classification problem. 
In particular, given a query budget $M=|\Nlb|$, find the set of data items $\Nlb\in\mathcal N$, whose labels' knowledge minimizes the  conditional 
risk in the classification of the remaining data items $\Nul \triangleq \mathcal N\backslash\Nlb$.  
Our analysis will not be limited to a specific loss function, but will be applicable to arbitrary loss functions, among which popular examples include the norm functions, (i.e.,  $\mathcal{C}(\hat{\Lbl},\Lbl)=\| \hat{\Lbl} -\Lbl \|_p^p$ (with $p=\{1,2\}$ being the most commonly adopted choices) and the binary loss function,  i.e.,  $\mathcal{C}(\hat{\Lbl},\Lbl)=1\{\hat{\Lbl}\neq \Lbl\}$. 

 \begin{table*}[h]
 \begin{align}
& \text{Define:}\,\,\, \Lbl_{m,m'} = 
\begin{cases}
\big[ \Lbl_{\Nlb^{(m-1)}}, L_{n_m}= \ell_{n_m},\dots,L_{n_{m'}}=\ell_{n_{m'}} \big] , \quad m'\geq m \notag\\
\Lbl_{\Nlb^{(m-1)}}  , \quad m' = m-1
\end{cases} \\
& \text{Initialize:}\,\,\, \mathcal{J}^{(M)}\Big( L_{n_{M}}, \Lbl_{m,M-1}, \Edg \Big) = \mathcal{R}\Big(\hat{\Lbl}_{\Nul^{(m)}}^{*} | L_{n_{M}}, \Lbl_{m,M-1}, \Edg \Big),  \,  \forall  {n_{M}} \in {\Nul^{(m)}},  
\qquad \qquad \qquad  \forall  L_{n_{i}} \in \mathcal L, \, \,   i=m,\ldots, M-1  \label{Eq:Initial_Point} \\
& \text{for:}\,\,\, m'=M-1,\ldots, m:
\quad \mathcal{J}^{(m')}\Big( L_{n_{m'}}, \Lbl_{m,m'-1}, \Edg \Big) = \underset{n_{m'+1}\in\Nset\backslash\{n_1\cup\cdots\cup n_{m'}\}}{\mathrm{min}} \,\,\mathbb{E}\Big[\mathcal{J}^{(m'+1)}\Big(  L_{n_{m'+1}}, \Lbl_{m,m'}, \Edg \Big)\Big]  \label{Eq:Recursive} \\
& \text{Return:}\,\,\, \mathcal{J}^{(m)}\big( L_{n_m}, \Lbl_{m,m-1}, \Edg \big) \notag
\end{align}	
\hrulefill
\end{table*}

\section{Optimal Active Learning 
Policy}
\label{Optimal Active Learning (query selection) Policy}

We now focus on the process of selecting $M$ items to query for labels such that the conditional risk associated with the classification of the remaining $N-M$ items 
is minimized. 
Intuition suggests that (i) the optimal active learning policy should be implemented in an iterative fashion, where at each iteration one new node to be queried is chosen based on a criterion that takes into account previously revealed information;
(ii) the query choice at the $m$-th iteration should account for all possible expected \textit{future} queries at iterations $m+1,\dots,M$.
To this end, we  introduce the {\em label query vector} $\ell_{m,m'}$, defined as the set of labels already revealed up to iteration $m$ and a possible realization of future queries up to iteration $m'$.

The following theorem formally describes the optimal query selection policy.

\begin{theo}\label{Theo:OptQuery}
Given a collection of items $\mathcal N$ with unknown labels $\Lbl_{\mathcal N}$, 
a realization of the observable $\Edg$, 
and a total query budget $M$, the 
{\em optimal} (in the sense of minimizing the conditional risk) set of items to be queried for labels,  is given by $\Nlb^{*}=\Nlb^{(M)}$ and is obtained according to the following \textit{iterative procedure}:
	\begin{IEEEeqnarray}{rll}\label{Eq:Nupdate} 
		\Nlb^{(m)} = \begin{cases}
			\emptyset,\quad &\mathrm{if} \,\,m=0\\
			\Nlb^{(m-1)}\cup n_{m}^*, &\mathrm{otherwise},
		\end{cases}
	\end{IEEEeqnarray}
	where $n_{m}^*$ denotes the { optimal} item to be queried at the $m$-th iteration. 
	In \eqref{Eq:Nupdate},  $n_{m}^*$ is given by 
    \begin{IEEEeqnarray}{lll}\label{Eq:nOptgeneral} 
		n_m^*=\underset{n_m\in{ \Nul^{(m)}}}{\mathrm{argmin}} \,\,\mathbb{E}\Big[ \mathcal{J}^{(m)}\big(L_{n_m}, \Lbl_{\Nlb^{(m-1)}}, \Edg \big)  \Big],
	\end{IEEEeqnarray}
where ${\Nul^{(m)}} \triangleq \Nset \, \backslash{\Nlb^{(m-1)}}$  is the set of unlabeled items at iteration $m$, and $\mathcal{J}^{(m)}\big(L_{n_m}, \Lbl_{\Nlb^{(m-1)}}, \Edg \big)$ is obtained via the recursive procedure described in   \eqref{Eq:Initial_Point}-\eqref{Eq:Recursive}.

\end{theo}
\begin{IEEEproof}
The proof is given in Appendix~\ref{App:Theo_OptQuery}.
\end{IEEEproof}


\begin{remk}
We note that $\mathcal{J}^{(m)}\big(L_{n_m}, \Lbl_{\Nlb^{(m-1)}}, \Edg \big)$ 
  is indicative of 
  the minimum conditional risk that would be incurred at the end of query process for the classification of the $N-M$ unlabeled items when the $m$-th queried item is $n_m$ with random label $L_{n_m}$, given  the observation $\Edg$ and the labels of the previous $m-1$ queried items $\Lbl_{\Nlb^{(m-1)}}$, and averaged over the labels associated with the possible choices taken in the next $M-m$ iterations. 
  \end{remk}

 \begin{figure*}
	\begin{minipage}{0.49\linewidth}
		\centering
		\includegraphics[width=1\linewidth]{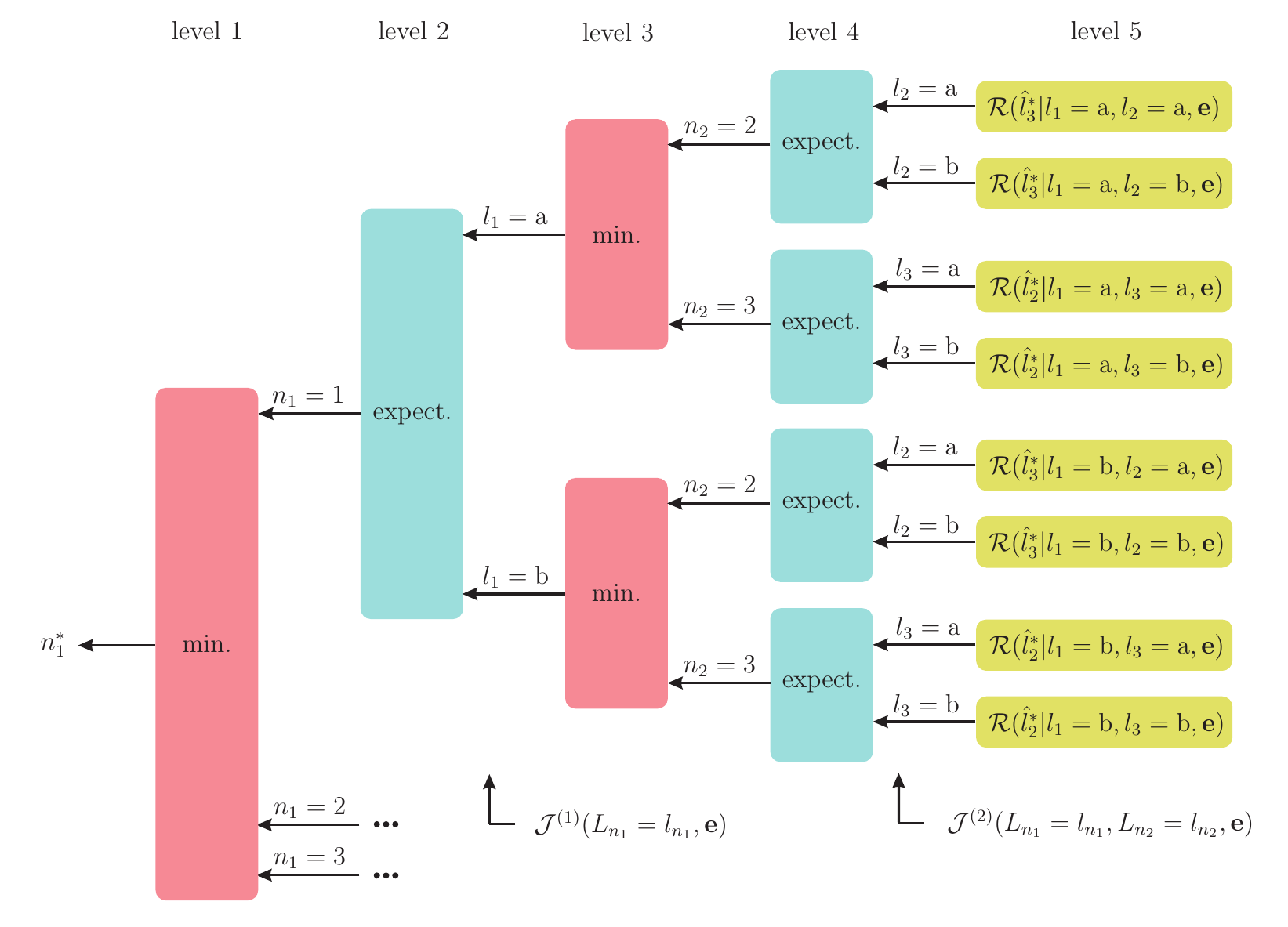}
			\caption{ Illustration of the operations involved in the iterative query selection procedure 
			\eqref{Eq:nOptgeneral} for the case of $\Nset=\{1,2,3\}$,  $\Lset=\{\mathrm{a},\mathrm{b}\}$, and $M=2$.
		\vspace{-0.3cm}}
		\label{Fig:TreeOptimal}
	\end{minipage}
\begin{minipage}{0.02\linewidth}
	\quad
\end{minipage}
	\begin{minipage}{0.49\linewidth}
		\centering
		\includegraphics[width=1\linewidth]{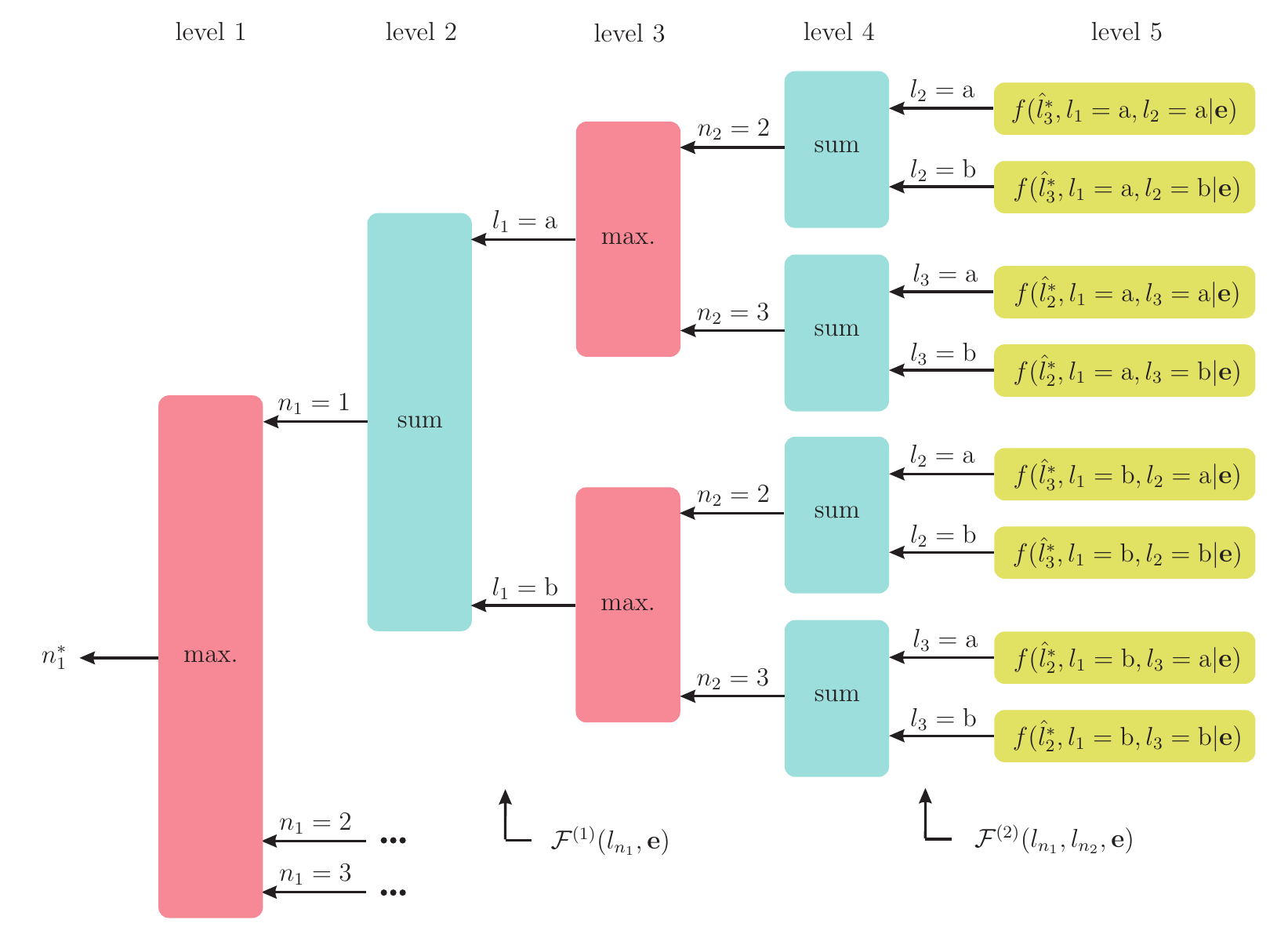}
			\caption{ Illustration of the operations involved in the iterative query selection procedure 
			\eqref{Eq:nOptgeneralMAP},  
			for the case of $\Nset=\{1,2,3\}$,  $\Lset=\{\mathrm{a},\mathrm{b}\}$, and $M=2$. \vspace{-0.3cm}}
		\label{Fig:TreeOptimalMAP}
	\end{minipage}
\end{figure*}

The overall iterative procedure in \eqref{Eq:Nupdate}-\eqref{Eq:nOptgeneral} can be illustrated via a tree structure, 
as shown in Fig.~\ref{Fig:TreeOptimal} for a simple example with $N=3$ items, $|\mathcal L|=2$ labels, and $M=2$ queries.  
In general, the {\em query selection tree} is composed of $2M+1$ levels, where the set of incoming edges at level $2i-1, \,\,i=1,\dots,M$, corresponds to all possible identities of the $i$-th queried item, and the set of incoming edges at level $2i, \,\,i=1,\dots,M$, corresponds to all possible associated labels. 
	Therefore, the reverse path from the root node to a given node at level $2i+1$ identifies 
	a specific set of item identities and associated labels for the first $i$ items to be queried. 
	The nodes in the tree are populated using the recursive procedure in  \eqref{Eq:Initial_Point}-\eqref{Eq:Recursive} starting at the leaf nodes. In particular, each leaf node is populated with  
	the minimum conditional risk that would be incurred if the sequence of queried items and associated labels were those
	given 
	by the path from the root node to the given leaf node (cf. (\ref{Eq:Initial_Point})). 
	Then, each node at level $2i, \,\,i=1,\dots,M,$ is populated by performing the \textit{expectation},  i.e., the weighted sum 
	of the quantities passed by its descendants, whereas each node at level $2i-1, \,\,i=1,\dots,M,$ is populated by performing the \textit{minimization} of the quantities passed by its descendants (cf. \eqref{Eq:Recursive}). 
	  Hence, the value stored in each node at level $2i-1, \,\,i=1,\dots,M,$ is 
	 	indicative of 
	 the minimum conditional risk that would be incurred at the end of query process upon the classification of the $N-M$ unlabeled items for each possible  $i$-th queried item, given that the label of the $i$-th queried item and the identities and labels of the previous $i-1$ queried items are those indicated by the path from that node to the root of the tree, and averaged over the identities and labels associated with all possible queried items in the next $M-i$ iterations, represented by the subtree rooted at the given node.
	At the beginning of the iterative procedure ($m=1$), we start at the root node and choose the identity of the first item to be queried by selecting the incoming edge with minimum incoming node value. After querying and obtaining the label of the chosen item, we move along the associated edge to the corresponding node at level $3$, and start the next iteration. The iterative procedure follows in this fashion until reaching a leaf node, at which point all items in $\Nlb^{*}$ have been selected and their associated labels $\Lbl_{\Nlb^{*}}$ revealed.

\section{\GREEN Information-Theoretic Bounds} 

{\GREEN In this section, 
we leverage recent results that relate the probability of error of the classical maximum-a-posteriori (MAP) classifier with the R\'{e}nyi entropy \cite{Sason_Verdu_MHyp_2018} to 
derive tight information-theoretic bounds on the  active learning cost-performance tradeoff.}
To this end, we first particularize Theorem~\ref{Theo:OptQuery} to the case of the binary loss function, under which the optimal classifier is the MAP classifier, and then
derive upper and lower bounds on the cost-performance tradeoff dictated by the probability of correct classification of active learning based semi-supervised MAP classification as a function of the query budget $M$. 

\subsection{\GREEN MAP Classification} 

In the following corollary, we particularize Theorem~\ref{Theo:OptQuery} to the case of the binary loss function. 
In this case, computing the minimum conditional risk at a given  leaf node of the query selection tree can be replaced by computing the maximum 
conditional posterior probability,  $f_{\mathbf{L}|\mathbf{E}}({\Lbl}_{\Nul}
|  \Llb, \Edg )$. 
To see this, consider the notation introduced in Theorem~\ref{Theo:OptQuery} and let $\Lbl_{1,M} = \big[ L_{n_1}= \ell_{n_1},\dots,L_{n_{M}}=\ell_{n_{M}} \big]$ denote the possible sequence of queried items' labels associated with a given leaf node in the query selection tree, and  $\mathcal{R}\big(\hat{\Lbl}_{\Nul}^{*} |  \Lbl_{1,M}, \Edg \big)$ the associated  minimum conditional risk. {\GREEN Then, $\mathcal{R}\big(\hat{\Lbl}_{\Nul}^{*} |  \Lbl_{1,M}, \Edg \big)$  is equal to  $1- f(\hat{\Lbl}_{\Nul}^* | \Lbl_{1,M}, \Edg)$, whereby $\hat{\Lbl}_{\Nul}^*$ can be  equivalently obtained by maximizing $f(\hat{\Lbl}_{\Nul}, \Lbl_{1,M}| \Edg)$.}

\begin{corol}\label{Corol:OptQueryMAP}
	For the binary loss function, the optimal query selection rule in \eqref{Eq:nOptgeneral} simplifies~to 
	\begin{IEEEeqnarray}{lll}\label{Eq:nOptgeneralMAP} 
		n_m^*=\underset{n_m\in{ \Nul^{(m)}}}{\mathrm{argmax}} \,\,\sum_{l_{n_m}\in\Lset} \mathcal{F}^{(m)}\big(\ell_{n_m},\Lbl_{\Nlb^{(m-1)}},\Edg \big),
	\end{IEEEeqnarray}
	where $\mathcal{F}^{(m)}\big(\ell_{n_m},\Lbl_{\Nlb^{(m-1)}},\Edg \big)$ is obtained via the recursive procedure given in
	 \eqref{Eq:Initial_Point}-\eqref{Eq:Recursive} with \eqref{Eq:Initial_Point} replaced by
	 $\mathcal{F}^{(M)}\big(\ell_{n_{M}}, \Lbl_{m,M-1}, \Edg \big) = f\big(\hat{\Lbl}_{\Nul^{(m)}}^{*},  \ell_{n_{M}}, \Lbl_{m,M-1}| \Edg \big) $ 
	 and \eqref{Eq:Recursive} replaced by 
	{\GREEN	 
	 \begin{IEEEeqnarray}{lll} 
	 \mathcal{F}^{(m')}\Big(\ell_{n_{m'}},\Lbl_{m,m'-1}, \Edg\Big)  \nonumber \\
	 =\!\! \underset{n_{m'+1}\in\Nset\backslash\{n_1\cup\cdots\cup n_{m'}\}}{\mathrm{max}} \,\,\sum_{\ell_{n_{m'+1}}\in\Lset} \!\! \mathcal{F}^{(m'+1)}\Big(\ell_{n_{m'+1}}, \Lbl_{m,m'},  \Edg \Big) \nonumber.
	 \end{IEEEeqnarray}
} 

	
Analogous to the 
role of $\mathcal{J}^{(m)}\big(L_{n_m}, \Lbl_{\Nlb^{(m-1)}}, \Edg \big)$ in Theorem~\ref{Theo:OptQuery}, here $\mathcal{F}^{(m)}\big(\ell_{n_m}, \Lbl_{\Nlb^{(m-1)}}, \Edg \big)$ 
is indicative of the probability of correct classification that would be obtained at the end of query process for the classification of the $N-M$ unlabeled items when the $m$-th queried item is $n_m$ with random label $L_{n_m}$, given the observation $\Edg$ and the labels of the previous $m-1$ queried items $\Lbl_{\Nlb^{(m-1)}}$, and averaged over the item labels associated with the possible choices taken in the next $M-m$ iterations. {\GREEN Moreover, similar to Theorem~\ref{Theo:OptQuery}, the iterative procedure in Corollary~\ref{Corol:OptQueryMAP} can be illustrated by the graph in Fig.~\ref{Fig:TreeOptimalMAP}, where compared to Fig.~\ref{Fig:TreeOptimal}, the minimization and expectation blocks and the conditional risk $\mathcal{R}\big (\hat{\Lbl}^*_{\Nul}|   \Llb, \Edg  \big)$ are replaced by the maximization and summation blocks and the posterior probability $f(\hat{\Lbl}^*_{\Nul},\Llb|\Edg  \big)$, respectively.} 
\end{corol}


\begin{figure*}
	\begin{minipage}{0.33\linewidth}
		\centering
		\includegraphics[width=1\linewidth]{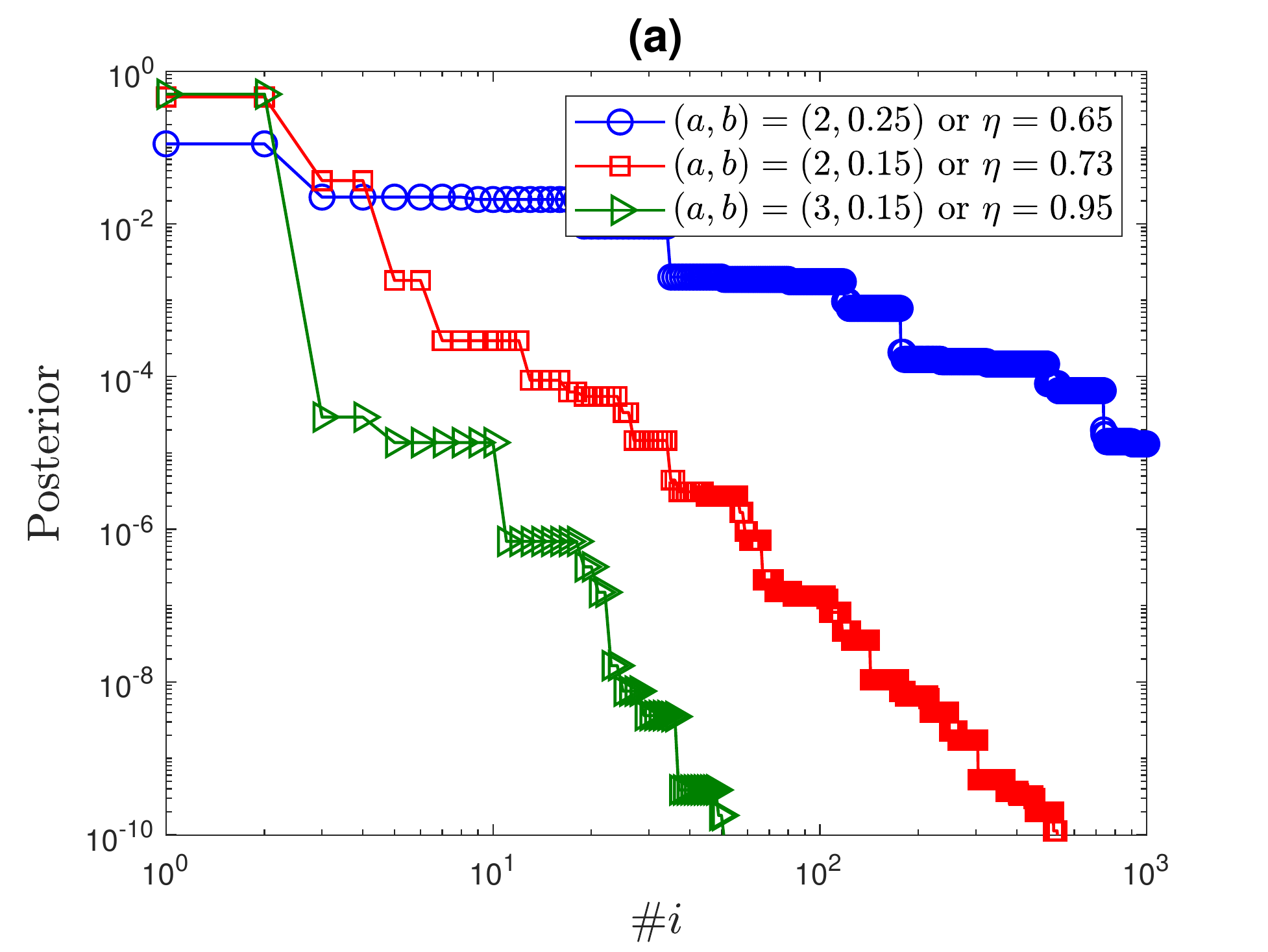}
	\end{minipage}
	\begin{minipage}{0.33\linewidth}
		\centering
		\includegraphics[width=1\linewidth]{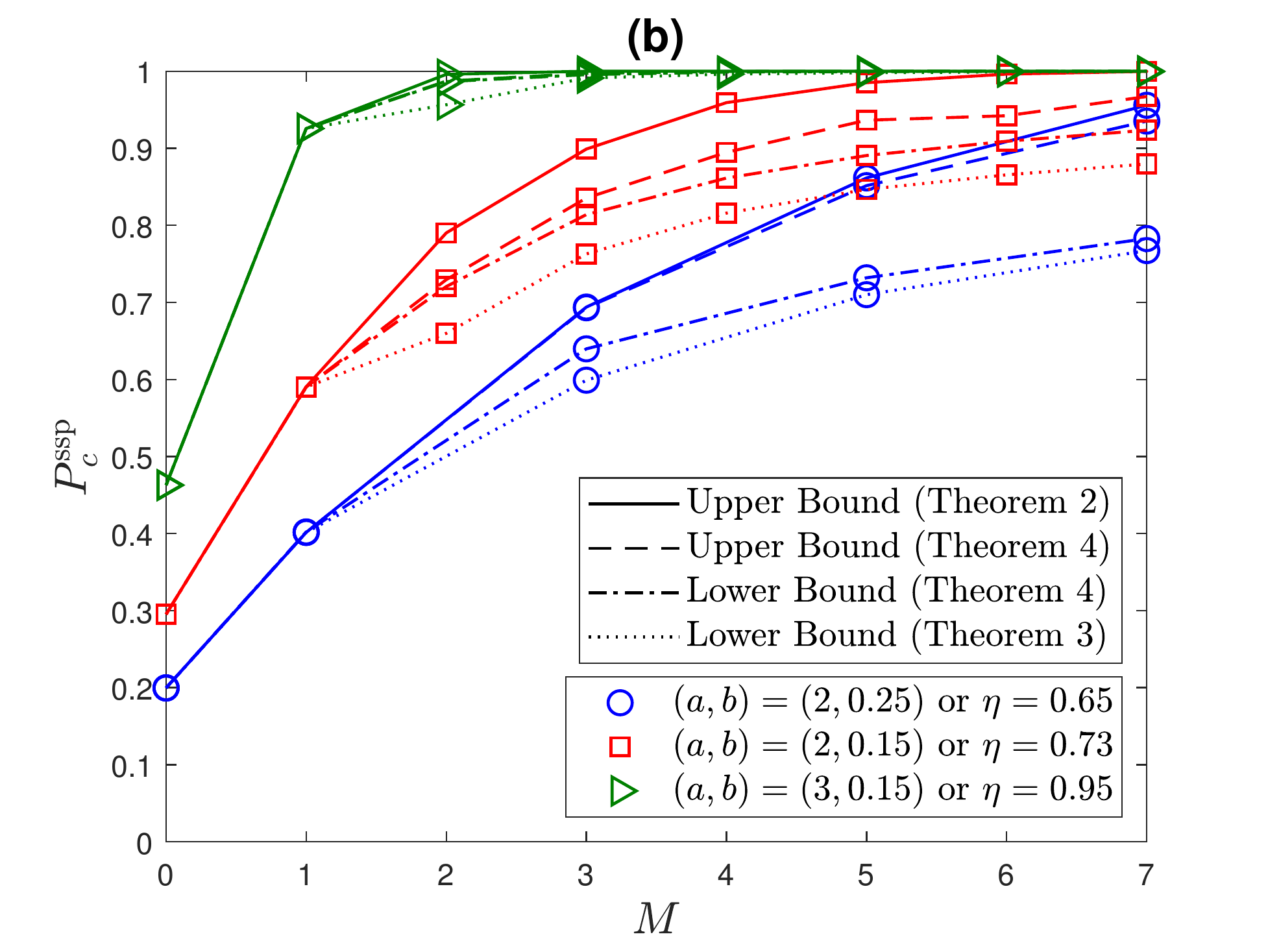}
	\end{minipage}
	\begin{minipage}{0.33\linewidth}
		\centering
		\includegraphics[width=1\linewidth]{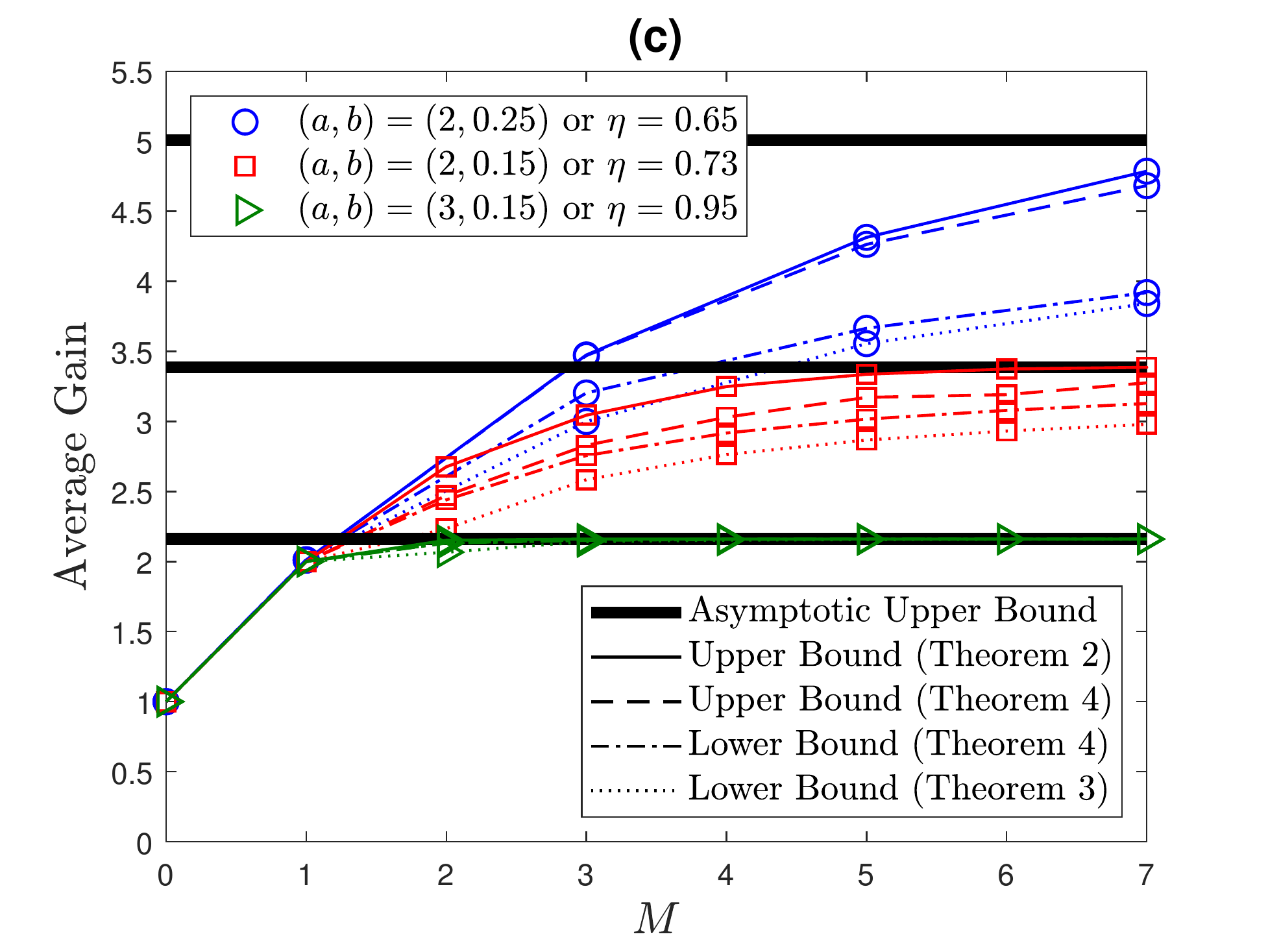}
	\end{minipage}
	\caption{(a) Posterior probabilities ordered in descending order. 
	(b) Probability of correct classification vs. query budget. (c) Average active learning gain. }
	\vspace{-0.3cm}
	\label{Fig:Sim}
\end{figure*}

\subsection{Active Learning Cost-Performance Tradeoff}



Let us first introduce the following R\'{e}nyi Entropy definitions.

\begin{defin}[\hspace{-0.1mm}\cite{Sason_Verdu_MHyp_2018}]
	Let $p_X$ denote the probability mass function of random variable $X$ taking values on  a discrete set $\mathcal X$. 
	The R\'{e}nyi entropy of order $\alpha \in (0,1) \cup (1,\infty)$  of $X$, denoted by $H_\alpha(X)$ is defined as
	\begin{eqnarray}
	H_\alpha(X)= \frac{1}{1-\alpha} \log \sum_{x\in \mathcal X} p_x^\alpha(x). 
	\end{eqnarray}
\end{defin}

\begin{defin}[\hspace{-0.1mm}\cite{Sason_Verdu_MHyp_2018}]
	Let $p_{XY}$  denote the joint probability mass function defined on $\mathcal X \times \mathcal Y$, where $X$ and $Y$ are discrete random variables. The Arimoto-R\'{e}nyi conditional entropy of order $\alpha \in (0,1) \cup (1,\infty)$ of $X$ given $Y$ is defined as 
	\begin{IEEEeqnarray}{lll}\label{conditional0}
		\scriptsize 
		H_\alpha(X|Y)= \frac{\alpha}{1-\alpha} \log \sum_{y\in \mathcal Y} 
		p_Y \exp{\left \{\frac{1-\alpha }{\alpha } H_\alpha(X|Y=y) \right\}} , 
		\nonumber 
	\end{IEEEeqnarray}
	where
	\begin{eqnarray}
	\displaystyle H_\alpha(X|Y=y)= \frac{\RED 1}{1-\alpha} \log \sum_{x\in \mathcal X} 
	p_{X|Y}^\alpha(x|y).
	\label{conditional1}
	\end{eqnarray}
\end{defin}

Letting $P_c^{\pi,\mathrm{ssp}}(\Edg)$ and $P_c^{\mathrm{opt},\mathrm{ssp}}(\Edg)$  denote the conditional (on the observation $\Edg$) correct classification probabilities of the semi-supervised {MAP} classifier under arbitrary query policy $\pi$ and under the optimal query policy of Corollary~\ref{Corol:OptQueryMAP}, respectively, the following theorems follow.

\begin{theo}
\label{Renyi_upperbound}
Under 
an arbitrary query selection policy $\pi$ with query budget $M$,
the  correct classification conditional probability $P_c^{\pi,\mathrm{ssp}}(\Edg)$  
can be upper bounded as 
\begin{eqnarray}
\log \left (P_c^{\pi,\mathrm{ssp}}(\Edg) \right)  
\leq  \!\!\! \! 
\inf_{\alpha \in (-\infty,\!  -1)}  \! 
\frac{1}{\alpha} \!  \left( H_{\frac{\alpha}{\alpha+1}} ({\bf L}|{\bf E}=\Edg)- \!  {M} \log(|\Lset|) \right)
\nonumber 
\end{eqnarray}

\end{theo}
\begin{IEEEproof}
	{\RED The proof is given in Appendix B.}
\end{IEEEproof}

{\RED 
\begin{defin}\label{Def:PermInv}
The posterior probabilities $f(\Lbl|\Edg)$ are referred to as label permutation invariant if  
for any label configuration $\Lbl$, 
permuting the label of every node in one class with the label of another class yields the same posterior probability.
\end{defin}
}

\begin{theo}
\label{Renyi_lowerbound}
{\RED  Let  $\gamma= |\Lset|!(M-|\Lset|+2)$ if $ M \geq |\Lset| -1$, and $\gamma =\frac{ |\Lset|!}{( |\Lset|-M)!} $  otherwise.
}
Under the optimal query selection policy of Corollary~\ref{Corol:OptQueryMAP} with query budget $M$,
and {\RED label permutation invariant posteriors},
the correct classification conditional  probability $P_c^{\mathrm{opt}, \mathrm{ssp}}(\Edg)$
can be lower bounded as  $P_c^{\mathrm{opt}, \mathrm{ssp}}(\Edg)  \geq   \sum_{m={\RED 1}}^{\gamma}
[\mathcal{B}^{(m)}]^+$, where 
\begin{IEEEeqnarray}{lll}
\mathcal{B}^{(m)} = & 
\Big[1-   
\!\!\! \inf_{k  \in \mathbb{S}, \, \alpha \in (1,\infty) } 
\! \!\!  
\frac{  \mathcal{H}_\alpha^{(m)} - 
k^{\frac{1}{\alpha}+1} +(k-1)(k+1)^{\frac{1}{\alpha}} }
{ k(k+1)^{\frac{1}{\alpha}} - k^{\frac{1}{\alpha}}(k+1)
}\Big]\nonumber\\ 
&\quad \Big(1- \sum_{j={\RED 1}}^{m-1}  \mathcal{B}^{(j)}\Big),
\nonumber
\end{IEEEeqnarray}
with $\mathcal{H}_\alpha^{(m)}= \exp{ \big\{ \frac{1-\alpha}{\alpha}  H^{(m)}_\alpha({\bf L}|{\bf E}=\Edg) \big\} }$, 
\begin{IEEEeqnarray}{lll}
H^{(m)}_\alpha({\bf L}|{\bf E}=\Edg) =  &{\RED \frac{1}{1-\alpha}} \Big[\log   \Big( \exp{\big\{{\RED(1-\alpha)}H_\alpha({\bf L}|{\bf E}=\Edg )}\big\}  \nonumber  \\
& - \Delta_\alpha^{(m)} \Big)  
- \alpha \log   \Big(1- \sum_{j={\RED 1}}^{m-1}  \mathcal{B}^{(j)}\Big) \Big], 
\label{iteration}
\end{IEEEeqnarray}
$\Delta_\alpha^{(m)} \! \!\! =  \!\! \sum_{j={\RED 1}}^{m-1} \!  \! \left( \mathcal{B}^{(j)}\right)^\alpha \! \! \! $, and
$\mathbb{S}  \!   = \! \!   \{ k  \!  \in \!  \mathbb{N} \! \!  :   \log(k) \! \!  \leq \!  H^{(m)}_\alpha \! ({\bf L}|{\bf E} \! = \! \Edg) \!  <  \log(k+1)\}$.
\end{theo}
\begin{IEEEproof}
	{\RED The proof is given in Appendix C.}
	\end{IEEEproof}
%
\begin{remk}
{\RED Letting $\gamma=M+1$, the lower bound in Theorem~\ref{Renyi_lowerbound} holds for arbitrary posteriors.}
\end{remk}
\begin{remk}
{\RED In Section~\ref{Sec:Sim}, we show that under the  SBM, which exhibits  
permutation invariant posteriors, the lower bound in Theorem \ref{Renyi_lowerbound} is tight. }
\end{remk}

It is important to know that the above R\'enyi Entropy based bounds are especially useful for large-scale learning settings due to their computation scalability. 
In the following, we derive alternative bounds that exploit the structure of the optimal query selection policy to provide slightly tighter bounds at the expense of increased exponential complexity in the number of data items $N$, and hence suited for 
small-scale data sets. 
We state these bounds in terms of 
the following definitions. 

\begin{defin}\label{gain}
We define the {\em active learning gain} of query selection policy $\pi$ under observation $\Edg$ as 
$\Upsilon^{\pi}(\Edg) = \frac{P_c^{\pi,\mathrm{ssp}}(\Edg)}{P_c^{\mathrm{usp}}(\Edg)}$, 
where $P_c^{\mathrm{usp}}(\Edg)$ 
denotes the conditional correct classification probability 
of the unsupervised {MAP} classifier. 
\end{defin}

\begin{defin}\label{order}
We define the {\em normalized accuracy} of label estimate $\Lbl$ as $\Phi(\Lbl|\Edg)=\frac{f(\Lbl | \Edg)}{f(\Lbl^* | \Edg)}$, where $\Lbl^*= {\mathrm{argmax}}_{\Lbl}\,\,f(\Lbl|\Edg)$ denotes the unsupervised {MAP} estimate. 
We then use $\Lset^{N} =\{\Lbl_{(1)}, \Lbl_{(2)}, \ldots, \Lbl_{(|\Lset|^{N})}\}$ to denote the  set of all possible label estimates ordered according to their normalized accuracy  (i.e., posterior probability), where $\Phi(\Lbl_{(1)}|\Edg)\geq \cdots\geq \Phi(\Lbl_{(|\Lset|^{N})})|\Edg)$. 
\end{defin}




\begin{theo}\label{Theo:LowerMGaingeneral}
Under the optimal query selection policy of Corollary~\ref{Corol:OptQueryMAP} with query budget $M$, the active learning gain can be upper and lower bounded as
	\begin{IEEEeqnarray}{lll}\label{Eq:BoundSupUnsupEdg1}
 \sum_{i=1}^{\Gamma}  \Phi(\Lbl_{(i)}|\Edg) \leq  \Upsilon^{\mathrm{opt}}(\Edg) \leq \sum_{i=1}^{|\Lset|^{M}} \Phi(\Lbl_{(i)}|\Edg), 
	\end{IEEEeqnarray}
 where $\Gamma$ is the largest  index of the ordered label estimates 
such that there exist
 $M$ items whose label configurations corresponding to the first $\Gamma$ posterior probabilities are distinct.
\end{theo} 

\begin{IEEEproof}
	The proof is given in Appendix~\ref{App:TheoLowerMGaingeneral}.
\end{IEEEproof}

\begin{remk}\label{Remk:Perm}
 The previous lower bound can be stated more explicitly
if we assume that the posterior probability 
$f(\Lbl|\Edg)$ is {\em label permutation invariant}, see Definition~\ref{Def:PermInv}. 
 %
Under this assumption, $\Gamma$ can be lower bounded as 
$\Gamma \geq |\Lset|!(M-|\Lset|+2)$ if $ M \geq |\Lset| -1$, and as $\Gamma \geq\frac{ |\Lset|!}{( |\Lset|-M)!} $ otherwise.
\end{remk}

\section{Simulation Results}\label{Sec:Sim}

We now present simulation results in the context of the SBM, whose model is given in \eqref{functionlink0}, with $N=15$ nodes, and two communities with identical membership probabilities, $q_{\mathrm{in}}=\frac{a\log(N)}{N}$, and $q_{\mathrm{out}}=\frac{b\log(N)}{N}$. 
Recall that for the planted SBM, i.e., when the sizes of communities are identical, exact clustering is asymptotically (as $N\to\infty$) possible if and only if $\eta=\frac{\sqrt{a}-\sqrt{b}}{\sqrt{2}}>1$ \cite{abbe2016exact}. Here, we focus on the interesting and challenging regime of finite $N$ and $\eta<1$. 
In Fig.~\ref{Fig:Sim}a, we show one realization of the ordered posterior probabilities for three statistical scenarios, namely Scenario~1: $(a,b)=(2,0.25)$, Scenario~2:  $(a,b)=(2,0.15)$, and Scenario~3:  $(a,b)=(3,0.15)$. 
Observe 
that the smaller the value of $\eta$, the larger the number of label configurations with non-negligible posterior, which implies that a larger number of queries is required 
to ensure a given probability of correct classification. This is shown in Fig.~\ref{Fig:Sim}b, where we plot the bounds on $P_c^{\mathrm{ssp}}=\mathbb{E}[P_c^{\mathrm{ssp}}(\mathbf{E})]$ of the MAP classifier averaged over $50$ graph realizations vs. the number of queries $M$ for the aforementioned three scenarios. As expected, $P_c^{\mathrm{ssp}}$ ultimately approaches one with sufficient number of queries; however, the required number of queries depends on the data statistics. For example, to achieve $P_c^{\mathrm{ssp}}>0.75$, we require $M\geq 1$, $3$, and $7$ in Scenarios 1, 2, and 3, respectively. 
Fig.~\ref{Fig:Sim}b also shows that the R\'{e}nyi entropy bounds of Theorems~\ref{Renyi_upperbound} and \ref{Renyi_lowerbound}  are not too far from to their respective bounds of Theorem~\ref{Theo:LowerMGaingeneral}, which supports their usefulness, given their computation scalability advantage, as they avoid the exponential complexity required to compute the ordered posteriors in Theorem~\ref{Theo:LowerMGaingeneral}.
Finally, Fig.~\ref{Fig:Sim}c plots the average active learning gain. 
 The black solid lines indicate the trivial upper bound 
 on the average active learning gain,   $\frac{1}{f(\Lbl^*|\Edg)}$ averaged over $\Edg$, achieved in the asymptotic regime when the number of queries is sufficient for perfect classification.
Note again how the R\'enyi Entropy bounds are not too far from those of Theorem 4, and significantly tight, especially for $\eta$ close to 1. 
Observe how the gain increases as $\eta$ decreases, which is the regime where active learning is most beneficial.


\appendices

\section{Proof of Theorem~\ref{Theo:OptQuery}}\label{App:Theo_OptQuery}

{\RED Notation: For ease of exposition, we use $\mathbb E[g(X,Y)|Y=y]$ and $\mathbb E[g(X,y)]$ interchangeably.}

For a given  {\RED observation} $\bf E=\Edg$, 
an arbitrary query selection policy $\pi$ can be described using the following iterative procedure:
\begin{align}\label{doctor0}
& n_m=\mathcal Q_\pi^{(m)} \left ( \Edg, \Lbl^{(\pi )}_{{\Nlb}^{\!\!(m-1)}} \right), m\in\{2,\dots, M\} \\
& n_1=\mathcal Q_\pi^{(1)} \left ( \Edg \right) \notag
\end{align}
where 
$\mathcal Q_{\pi}^{(m)}(\cdot)$, $m=\left\{1, \ldots, M \right\}$, denotes the (possibly random) rule established by policy $\pi$ for selecting the item to be queried at the $m$-th iteration, $n_m$, and 
$\Lbl^{(\pi )}_{{\Nlb}^{(m-1)}}$ denotes the set of items and associated labels revealed after the first $m-1$ iterations of policy $\pi$, 
with the convention of denoting the set of items and associated labels revealed after $M$ iterations by $\Lbl^{(\pi )}_{{\Nlb}}$.\footnote{{\RED Recall that we assume that the notation $L_{\mathcal A}$ and $\ell_{\mathcal A}$ indicates both the indices of items in $\mathcal A$ and its associated labels.}}
Note that {\RED the general iterative rule in \eqref{doctor0} assumes that} 
the value of  $n_m,\forall m$, may depend on 
\textit{previously} revealed labels, 
{\RED and hence includes as special cases the more standard batch and random policies. In particular, batch policies are equivalent to applying \eqref{doctor0} assuming that the choice of $n_m$ does not depend on the previously revealed labels,  
and random policies are equivalent to assuming $Q_{\pi}^{(m)}(\cdot)$ is a random function independent of the choices made in the previous iterations. }


%

For a given query selection policy $\pi$ with resulting label realization $\Llb^{(\pi )}$, let 
$\mathcal{R}\big (\hat{\Lbl}_{\Nul}^{*}|   \Llb^{(\pi )}, \Edg,  \big)$ denote the conditional risk (conditioned on  $\bf E= \Edg$ and $\bf L^{{\RED (\pi)}}_{\Nlb} = \Llb^{{\RED (\pi)}}$) obtained by the solution of the {\RED optimal} semi-supervised classifier in \eqref{Eq:OptClassification}, $\hat{\Lbl}_{\Nul}^{*}$. 

{\RED In the following, for ease of exposition, and unless specified otherwise, we will refer to the conditional risk under a given query policy as the conditional risk obtained by the optimal semi-supervised classifier in \eqref{Eq:OptClassification} under that given query policy.}

Then, the conditional risk (conditioned only on  $\bf E=\Edg$) 
under policy $\pi$, denoted by $\mathcal{R}_\pi(\Edg)$, is given by 
\begin{IEEEeqnarray}{lll}
	\mathcal{R}_\pi(\Edg) = \mathbb{E}
	\big[\mathcal{R}\big (\hat{\Lbl}_{\Nul}^{*}|  {\mathbf{L}^{(\pi )}_{\Nlb}}, \bf E \big)|  \bf E=\Edg \big].
	\label{Eq:AverageRiskPolicy}
\end{IEEEeqnarray}
Our goal is to identify the optimal query selection policy, denoted by $\pi^*$, that yields the smallest (over all possible query policies) 
conditional risk (conditioned on  $\bf E=\Edg$), i.e., 
\begin{IEEEeqnarray}{lll}
	\pi^*=\underset{\pi}{\mathrm{argmin}}\,\,\mathcal{R}_\pi(\Edg).
\end{IEEEeqnarray}
{\RED In line  with the notation of Theorem~\ref{Theo:OptQuery}, we directly use $\Lbl_{{\Nlb}^{(m)}}$  
to denote the set of items and associated labels revealed after the first $m$ iterations by the optimal policy $\pi^*$.}

 
To derive the optimal query selection policy, we follow an inductive argument. 
We first obtain the optimal policy for $M=1$ and $M=2$, and then generalize it to $M>2$.

\textbf{Case~1 ($M=1$):} The 
 conditional risk $\mathcal{R}_\pi(\Edg)$ under an arbitrary query policy $\pi$  is   given by 
\begin{IEEEeqnarray}{lll}
	\mathcal{R}_\pi(\Edg) = \mathbb{E}
	\big[   \mathcal{R}(\hat{\Lbl}_{\Nul}^{*}| L^{(\pi )}_{n_1}, \bf E )  \, \, |  \,
 \bf E=\Edg  \big] . 
  	\label{Eq:AverageRiskPolicy}
\end{IEEEeqnarray}
Therefore, the  conditional risk $\mathcal{R}_{\pi^*}(\Edg)$ under the optimal policy $\pi^*$ is given by 
\begin{IEEEeqnarray}{lll}
	\mathcal{R}_{\pi^*}(\Edg) = \underset{n_1\in\Nset}{\mathrm{min}}\,\,  \mathbb{E}
	\big[   \mathcal{R}(\hat{\Lbl}_{\Nul}^{*}| L_{n_1}, \bf E )  \, \, |  \,
 \bf E=\Edg  \big] . 
  	\label{Eq:AverageRiskPolicyyyy}
\end{IEEEeqnarray}
The optimal  policy $\pi^*$ will hence select item  $n_1^*$~as
 \begin{IEEEeqnarray}{lll}
 	n_1^*&=&\underset{n_1\in\Nset}{\mathrm{argmin}}\,\,
 \mathbb{E}
 \big[   \mathcal{R}(\hat{\Lbl}_{\Nul}^{*}| L_{n_1}, \bf E ) \,  \,| \,
 \bf E=\Edg  \big] \nonumber \\
	&=& \underset{n_1\in\Nset}{\mathrm{argmin}}\,\, 	\mathbb{E}
	\big[ \mathcal{J}^{(1)}\big( L_{n_1},  \Edg )  \big] , 
   	\label{Eq:rottan+1}
 \end{IEEEeqnarray}
where $\mathcal{J}^{(1)}\big( L_{n_1},  \Edg )$ follows the notation introduced in Theorem  \eqref{Theo:OptQuery} for $M=1$ and the 
expectation is with respect to the random label of item $n_1$ conditioned on $\Edg$. 


\textbf{Case~2 ($M=2$):} The  conditional risk $\mathcal{R}_\pi(\Edg)$ under an arbitrary query policy $\pi$  can be written as 
\begin{IEEEeqnarray}{lll}
	\mathcal{R}_\pi(\Edg) = \mathbb{E}
	\big[  \mathcal{J}_\pi^{(1)}\big( L^{(\pi )}_{n_1},  \bf E \big)  | \bf E=\Edg  \big],
	\label{Eq:AverageRiskPolicy00}
\end{IEEEeqnarray}
with
\begin{IEEEeqnarray}{lll}
\mathcal{J}_\pi^{(1)}\big( \ell^{(\pi )}_{n_1},  \Edg \big) 
&=& \mathbb{E}
 \big[  \mathcal{R}\big (\hat{\Lbl}_{\Nul}^{*}| L_{n_2}^{(\pi )}, { L}^{(\pi )}_{n_1} , {\bf E}  \big)
  | {\bf E}=\Edg, { L}^{(\pi )}_{n_1}=\ell^{(\pi )}_{n_1}  \big] 
\nonumber \\
&=& \mathbb{E}
 \big[  \mathcal{R}\big (\hat{\Lbl}_{\Nul}^{*}| L_{n_2}^{(\pi )}, \ell^{(\pi )}_{n_1} , \Edg \big] . 
\end{IEEEeqnarray}
The conditional risk under the optimal policy $\pi^*$ is then given by
\begin{IEEEeqnarray}{lll}
\mathcal{R}_{\pi^*}(\Edg)	
\!=\!\! \underset{n_1\in\Nset}{\mathrm{min}}\mathbb{E}
 \big[ 
\mathcal{J}^{(1)}\big( L_{n_1}, \bf E \big)  | \bf E=\Edg  \big]
\label{parac2}
\end{IEEEeqnarray}
with
\begin{IEEEeqnarray}{lll}
\mathcal{J}^{(1)} \! \big( \ell_{n_1} \!, \! \Edg \big)   &=& \!\!\! \!\!\!\!\underset{n_2\in\Nset\backslash\{n_1\} }{\mathrm{min}}\!\!\! \!\!\mathbb{E}
\big[ \mathcal{R}\big (\hat{\Lbl}_{\Nul}^{*}| L_{n_2}, L_{n_1}, {\bf E} \big) | {\bf E}\! =\!\Edg, L_{n_1} \!\!\!=\! \ell_{n_1}  \big]
\nonumber \\
&=& 
\!\!\!\!\underset{n_2\in\Nset\backslash\{n_1\} }{\mathrm{min}}\!\!\! \!\!\mathbb{E}
\ \big[ \mathcal{J}^{(2)}\big( L_{n_2}, \Lbl_{1,1}, \Edg \big) \big]
\label{parac1}
\end{IEEEeqnarray}
{\RED where $\Lbl_{1,1}=\ell_{n_1}$ follows from specializing the definition of $\Lbl_{m,m'}$ in Theorem 1 to $m=1,m'=1$. } 
Hence, the optimal  policy will select $n_1^*$ as 
\begin{IEEEeqnarray}{lll}\label{Eq:n1opt}
	n_1^*=\underset{n_1\in\Nset}{\mathrm{argmin}}\,\,\mathbb{E}
 \big[ 
\mathcal{J}^{(1)}\big( L_{n_1},  \Edg \big) \big]
\end{IEEEeqnarray}
and  $n_2^*$ as
\begin{IEEEeqnarray}{lll}\label{Eq:n2opt}
	n_2^*(\ell_{n^*_1})&=&
	\underset{n_2\in\Nset\backslash\{{n^*_1}\}}{\mathrm{argmin}}\,\,\mathbb{E}
	\big[  [\mathcal{R}\big (\hat{\Lbl}_{\Nul}^{*}| L_{n_2}, L_{n_1}, {\bf E} \big) | {\bf E}\! =\!\Edg, L_{n_1} \!\!\!=\! \ell_{n^*_1}  \big]
	\nonumber \\
	&=&
	\underset{n_2\in\Nset\backslash\{{n^*_1}\}}{\mathrm{argmin}}\,\,\mathbb{E}
	\big[\mathcal{J}^{(2)}\big( L_{n_2}, \Lbl_{2,1}, \Edg \big) \big] , \\
	&=&
	\underset{n_2\in\Nset\backslash\{{n^*_1}\}}{\mathrm{argmin}}\,\,\mathbb{E}
	\big[\mathcal{J}^{(2)}\big( L_{n_2},\Lbl_{\Nlb^{(1)}}, \Edg \big) \big] , 
	\label{parac2}
\end{IEEEeqnarray}
where from the definition of $\Lbl_{m,m'}$ in Theorem 1, we have that $\Lbl_{2,1} = \Lbl_{\Nlb^{(1)}}$, and since $\Lbl_{\Nlb^{(1)}}$ denotes the item label revealed up to iteration $1$ in the optimal policy, then $\Lbl_{\Nlb^{(1)}}=\ell_{n_1^*}$. 



To summarize, the optimal policy $\pi^*$ for $M=2$ 
requires: 1) computing $\mathcal{R}(\hat{\Lbl}_{\Nul}^{*}|\ell_{n_1},\ell_{n_2}, \Edg)$ for all $n_1$, $n_2\neq n_1$, and their possible label realizations; 2) computing $n_1^*$ from \eqref{Eq:n1opt} and asking the oracle for its label; and 3) given the revealed label $L_{n_1^*}=\ell_{n_1^*}$, computing $n_2^*$ from \eqref{Eq:n2opt} as $n_2^*=n_2^*(\ell_{n_1^*})$.

\textbf{Case~3 ($M>2$):} This case is a straightforward generalization of $M=2$. 
In fact, as for $M=2$, 
the conditional risk $\mathcal{R}_\pi(\Edg)$, achieved by a given query policy $\pi$,  can be written as 
\begin{IEEEeqnarray}{lll}
	\mathcal{R}_\pi(\Edg) = \mathbb{E}
	\big[  \mathcal{J}_\pi^{(1)}\big( L^{(\pi )}_{n_1},  \bf E \big)  | \bf E=\Edg  \big],
	\label{Eq:AverageRiskPolicy00}
\end{IEEEeqnarray}
with $ \mathcal{J}_\pi^{(1)}\big( \ell^{(\pi )}_{n_1},  \bf e \big) $ defined by the following recursion:
for all $m\in\{M-1,\dots,1\}$,  
\begin{IEEEeqnarray}{lll}
&&\mathcal{J}_\pi^{(m)}\big( \ell^{(\pi )}_{n_{m}}, {\bf \ell}^{(\pi )}_{\Nlb^{(m-1)}},  \Edg \big)
\nonumber \\
&&= \mathbb{E}
 \Big[ \mathcal{J}_\pi^{(m+1)}\big(  L^{(\pi )}_{n_{m+1}},  {\bf L}^{(\pi )}_{\Nlb^{(m)}},  {\bf E} ) \, \, \, |  \,
{\bf L}^{(\pi )}_{\Nlb^{(m)}}= \Lbl^{(\pi )}_{\Nlb^{(m)}}, {\bf E}=\Edg \Big]  \nonumber \\
&&
 \quad  \quad   \quad \quad  \quad  \quad  \quad = \mathbb{E}
 \Big[ \mathcal{J}_\pi^{(m+1)}\big(  L^{(\pi )}_{n_{m+1}},  {\bf \ell}^{(\pi )}_{\Nlb^{(m)}},  \Edg  \big) \Big],
\label{casso}
\end{IEEEeqnarray}
with  
$$
 \mathcal{J}_\pi^{(M)}\big(\ell^{(\pi )}_{n_{M}},\Lbl^{(\pi )}_{\Nlb^{(M-1)}}, \Edg \big)=\mathcal{R}\big (\hat{\Lbl}_{\Nul}^{*}| \ell^{(\pi )}_{n_{M}}, \Lbl^{(\pi )}_{\Nlb^{(M-1)}}, \Edg   \big) $$
and $ {\bf \ell}^{(\pi )}_{\Nlb^{(0)}}=\emptyset$,  $ {\bf \ell}^{(\pi )}_{\Nlb^{(m)}} = \{  \ell^{(\pi )}_{n_{m}},  {\bf \ell}^{(\pi )}_{\Nlb^{(m-1)}}\}.$ 

%
Consequently, the conditional risk obtained by the optimal policy, $\mathcal{R}_{\pi^*}(\Edg)$, can be found as a sequence of minimization and expectation operations. Specifically, 
we have that 
\begin{IEEEeqnarray}{lll}
	\mathcal{R}_{\pi^*}(\Edg) =  \underset{n_1\in \Nset }{\mathrm{min}}\mathbb{E}
	\big[  \mathcal{J}^{(1)}\big( L_{n_1},  \bf E \big)  | \bf E=\Edg ) \big],
	\label{Eq:AverageRiskPolicy00}
\end{IEEEeqnarray}
where, {\RED for $m'=M-1,\ldots, 1$, }
\begin{IEEEeqnarray}{lll}
&&\mathcal{J}^{(m')}\big( \ell_{n_{m'}},  \Lbl_{1,m'-1},  \Edg \big) =  
\nonumber \\
&&\underset{n_{m'+1}\in\Nset\backslash\{n_1\cup\cdots\cup n_{m'}\} }{\mathrm{min}} \mathbb{E}
\Big[  \mathcal{J}^{(m'+1)}\big(  L_{n_{m'+1}}, \Lbl_{1,m'},  \Edg  ) 
\Big],
\label{Eq:RiskGeneracasso}
\end{IEEEeqnarray}
with {\RED 
$$ \mathcal{J}^{(M)}\big(\ell_{n_{M}}, \Lbl_{1,M-1}, \Edg \big)=\mathcal{R}\big (\hat{\Lbl}_{\Nul}^{*}| \ell_{n_{M}},  \Lbl_{1,M-1}, \Edg   \big). 
$$ }

In summary, 
we have to first compute the conditional risk $\mathcal{R}(\hat{\Lbl}_{\Nul}^*|,\ell_{n_M},\dots,  \ell_{n_1}, \Edg)$ for all given vectors $[n_1,\dots,n_m]$,  $n_m\neq n_{m'},\forall m,m'$, and their possible labels. Subsequently, we compute the expectation with respect to $L_{n_M}$ (conditioned on ${\bf E} =\Edg $ and ${\bf L}_{\Nlb^{(M-1)}} = \Lbl_{\Nlb^{(M-1)}}$)  and then perform minimization with respect to $n_M$ as dictated by  \eqref{Eq:RiskGeneracasso}. This expectation-minimization operation continues for $L_{n_{M-1}}$, $L_{n_{M-2}}$, until the first item $L_{n_1}$. 

Starting from the expression of $\mathcal{R}_{\pi^*}(\Edg)$, it follows  immediately that the optimal policy will select 
$n_1^*$ as 
\begin{IEEEeqnarray}{lll}
n_1^* =  \underset{n_1\in \Nset }{\arg \mathrm{min}}\,\mathbb{E}
	\big[  \mathcal{J}^{(1)}\big( L_{n_1},  \bf E \big)  | \bf E=\Edg ) \big].
\end{IEEEeqnarray}

Furthermore,  for $m=2, \ldots, M$, 
$n_m^*$
will be selected  
 iteratively as 
 \begin{IEEEeqnarray}{lll} 
	n_m^*(\Lbl_{\Nlb^{(m-1)}}) =\underset{n_m\in{ \Nul^{(m)}}}{\mathrm{argmin}} \,\,\mathbb{E}_{L_{n_m}}\Big[ \mathcal{J}^{(m)}\big(L_{n_m}, \Lbl_{\Nlb^{(m-1)}}, \Edg \big)  \Big],
	\nonumber 
\end{IEEEeqnarray}
with $\Nul^{(m)}=\Nset\backslash\Nlb^{(m-1)}$, $\Nlb^{(m-1)}$ defined in \eqref{Eq:Nupdate} 
and, $  \mathcal{J}^{(m)}\big(L_{n_m}, \Lbl_{\Nlb^{(m-1)}}, \Edg \big)=  \mathcal{J}^{(m)}\Big( L_{n_{m'}}, \Lbl_{m,m-1}, \Edg \Big)  $ defined recursively by \eqref{Eq:Initial_Point}-\eqref{Eq:Recursive}.

 \begin{table*}[h]
 \begin{align}
&  \text{for:}\,\,\, m'=M-1,\ldots, m:  \nonumber \\ 
& \mathcal{J}^{(m')}\big(\ell_{n_{m'}}, \Lbl_{\Nlb^{(m-1)}},  \ell_{n_m},\dots,\ell_{n_{m'}}, \Edg \big) \nonumber \\
& =  \!\!\!\!\!\underset{n_{m'+1}\in\Nul^{(m')}\backslash\{n_{m'}\}}{\mathrm{min}}\,\,\mathbb{E} 
	 \big [ 
\mathcal{J}^{(m'+1)}\big(L_{n_{m'+1}},  {\bf L}_{\Nlb^{(m-1)}}, L_{n_m},\dots,L_{n_{m'}} {\bf E} \big) |  L_{n_m}= \ell_{n_m},\dots,L_{n_{m'}}=\ell_{n_{m'}}, 
{\bf L}_{\Nlb^{(m-1)}} = \Lbl_{\Nlb^{(m-1)}},  {\bf E} =\Edg \big]    \label{Eqrottotoo} 
\end{align}	
\hrulefill
\end{table*}

\section{Proof of Theorem~\ref{Renyi_upperbound}}\label{App:Renyi_upperbound}
In the following, we  prove the upper bound,  given in Theorem~\ref{Renyi_upperbound}, on the conditional (conditioned on $\Edg$)  probability of correct decision of the jointly optimal  semi-supervised MAP classifier and active learning policy, $P_c^{\mathrm{opt,ssp}}(\Edg)$. To do this end, let us recall \cite[Theorem 1]{Sason_VerduGeneral_2018}:
 
\begin{theo}
\label{th:Sason_VerduGeneral_2018}
Given a discrete random variable $X$ taking values on a set $\mathcal X$, a function $g: X \rightarrow (0,\infty)$, and
a scalar $\rho \neq 0$ then:
$$
\frac{1}{\rho} \mathbb{E}\Big[ g^\rho(X)\Big] \leq \!\!\! \inf_{\alpha \in (-\infty, -\rho)\setminus\{0\} }   \frac{1}{\alpha} \Big[ 
H_{\frac{\alpha}{\alpha+\rho}}(X) -\log \sum_{x \in \mathcal X} g^{-\alpha}(x)
\Big].
$$
\end{theo}
Before to use \cite[Theorem 1]{Sason_VerduGeneral_2018} to derive our proposed upper bound, let us recall that  for a given observable realization $\Edg$ and for a given realization $ \Lbl^{\mathrm{opt}}_{\Nlb}$,  the probability of correct classification of the semi-supervised MAP classifier conditioned on the augmented observable $(\Edg , \Lbl^{\mathrm{opt}}_{{\Nlb}})$ is 
$P_c^{\mathrm{opt,ssp}}(\Edg,\Lbl^{\mathrm{opt}}_{\Nlb})=  {\max}_{\Lbl_{\Nul}}\,\,f(\Lbl_{\Nul}|\Lbl^{\mathrm{opt}}_{\Nlb},\Edg)$. Therefore, 
the conditional (conditioned on $\Edg$) probability of correct classification of the jointly optimal semi-supervised MAP classifier and active learning is given by: 
\begin{IEEEeqnarray}{lll}
P_c^{\mathrm{opt,ssp}}(\Edg)\, &=\mathbb{E}\Big[ P_c^{\mathrm{opt,ssp}}({\bf E},{\bf L}^{\mathrm{opt}}_{\Nlb})  | {\bf E}= \Edg \Big] 
\label{doctor0optuff}
\end{IEEEeqnarray}
Note that since any arbitrary policy  without loss of generality can be described  using the iterative procedure  
in \eqref{doctor0}, the optimal policy can be illustrated as follow: 
\begin{eqnarray}
n^*_m=\mathcal Q_{\mathrm{opt}}^{(m)} \left ( \Edg,  \Lbl^{\mathrm{opt}}_{{\Nlb}^{\!\!(m-1)}} \right).
\label{doctor0opt}
\end{eqnarray}
where 
 $\mathcal Q_{\mathrm{opt}}^{(m)}(\cdot)$  with $m=\left\{1, \ldots, M \right\}$ indicates  the optimal query selection policy as described in  \eqref{Eq:nOptgeneralMAP} in Corollary~\ref{Corol:OptQueryMAP}.

Therefore, $n_{m}^*$ is a function of the 
observable realization $\Edg$, the previously optimally chosen data item indices ${\Nlb}^{\!\!(m-1)}$,  and their associated revealed label realization which we jointly denote by $\Lbl^{\mathrm{opt}}_{{\Nlb}^{\!\!(m-1)}}$.
Furthermore, from  Corollary~\ref{Corol:OptQueryMAP}, the optimal policy,  $\mathcal Q_{\mathrm{opt}}^{(m)}(\cdot)$,  is  a deterministic mapping which, 
 given its arguments, without loss of generality, we can assume  to return a unique data item index. In fact, if  multiple data items 
 satisfies \eqref {Eq:nOptgeneralMAP},  $\mathcal Q_{\mathrm{opt}}^{(m)}(\cdot)$ chooses the data item with e.g. the smallest index.

Conditioned on the observable $\bf E$, let $\mathcal P$ denote the set of all possible realizations of the two-component random vector, $\bf L_{\Nlb}$, resulting from the optimal query selection policy. Given that $\mathcal Q_{\mathrm{opt}}^{(m)}(\cdot)$ is iterative, deterministic, and returns a unique output, the cardinality of $\mathcal P$ is $|\Lset|^M$. Hence,  \eqref{doctor0optuff} admits the following expression:
\begin{IEEEeqnarray}{lll}
P_c^{\mathrm{opt,ssp}}(\Edg)\, &=\mathbb{E}\Big[ P_c^{\mathrm{opt,ssp}}({\bf E},{\bf L}^{\mathrm{opt}}_{\Nlb})  | {\bf E}= \Edg \Big]  \nonumber \\
&= \sum_{\Lbl_{\Nlb} \in \mathcal P  } P_c^{\mathrm{opt,ssp}}(\Edg,\Lbl_{\Nlb}) f(\Lbl_{\Nlb}|\Edg) \nonumber\\
&= \sum_{\Lbl_{\Nlb} \in \mathcal P  }  [  {\max}_{\Lbl_{\Nul}}\,\,f(\Lbl_{\Nul}|\Lbl_{\Nlb},\Edg) ]f(\Lbl_{\Nlb}|\Edg) \nonumber\\
&= \sum_{\Lbl_{\Nlb} \in\mathcal P }   {\max}_{\Lbl_{\Nul}}\,\,[f(\Lbl_{\Nul}|\Lbl_{\Nlb},\Edg) f(\Lbl_{\Nlb}|\Edg)] \nonumber \\
&= \sum_{\Lbl_{\Nlb} \in\mathcal P }   {\max}_{\Lbl_{\Nul}}\,\,f(\Lbl_{\Nul},\Lbl_{\Nlb}|\Edg). 
\label{triste0}
\end{IEEEeqnarray}
Therefore, $P_c^{\mathrm{opt,ssp}}(\Edg)$ is in general the sum of $|\Lset|^M$ posterior probabilities. Let now 
$\mathcal S^*$ be the set of all label configurations  associated to the $|\Lset|^M$ posterior probabilities in \eqref{triste0}. 
Furthermore, let $\Lset^{N} =\{\Lbl_{(1)}, \Lbl_{(2)}, \ldots, \Lbl_{(|\Lset|^{N})}\}$ denote the  set of all possible label estimates ordered according to their posterior probability  (i.e.,  normalized accuracy), as per Definition \ref{order} and  
let $X_{|\Edg}$  by the  mapping  that goes from $\Lset^{N} =\{\Lbl_{(1)}, \Lbl_{(2)}, \ldots, \Lbl_{(|\Lset|^{N})}\}$ to   $\{1, \dots, |\Lset|^{N}\}$  defined as  $X_{|\Edg}(\Lbl_{(i)})= i$. Therefore, $X_{|\Edg}$ is a random variable whose probability mass function  for all $i\in\{1, \dots, |\Lset|^{N}\}$, is  $ P(X_{|\Edg}=i)= f(\Lbl_{(i)}|\Edg)$.  Setting $g(i)$ such that:
\begin{IEEEeqnarray}{lll}
g(i)= 
\begin{cases} 
1, &  \mbox{if}   \,\,   \Lbl_{(i)} \in \mathcal{S}^* \\
0, & \mbox{otherwise}
\end{cases}
\end{IEEEeqnarray}
 and  applying , for $\rho=1$, Theorem \ref{th:Sason_VerduGeneral_2018}  to $g(X_{|\Edg})$, 
  Theorem~\ref{Renyi_upperbound} follows immediately.

\section{Proof of Theorem~\ref{Renyi_lowerbound}}\label{App:Renyi_lowerbound}
In the following, we  prove the lower bound,  given in  Theorem~\ref{Renyi_lowerbound}, on $P_c^{\mathrm{opt,ssp}}(\Edg)$.  
To do this we  first provide a lower bound on  the R\'{e}nyi entropy of a  random variable conditioned to a given event. 

Let $X$ and $Y$ two discrete random variables defined on $\mathcal X \times \mathcal Y$, and let $p(x|y )p(y)$ the associated  joint probability mass function.  Using  \cite[Eq (168)]{Sason_Verdu_MHyp_2018} in conjunction with \cite[Eq (170)]{Sason_Verdu_MHyp_2018}, 
it follows that the R\'{e}nyi entropy of $X$ conditioned to the event ${Y=y}$:
\begin{IEEEeqnarray}{lll}
H_\alpha(X | Y=y) \geq  
\frac{\alpha}{1-\alpha} \log \big (g_{ \alpha}(1-\max_{x\in \mathcal X} p(x|y ))\big)
\label{rottaca}
\end{IEEEeqnarray}
with 
$$
g_{ \alpha}(t)= \left(k(k+1)^{\frac{1}{\alpha}}-k^{\frac{1}{\alpha}}(k+1)\right)t +k^{\frac{1}{\alpha}+1} - (k-1)(k+1)^{\frac{1}{\alpha}}.
$$
Note that the right-hand side of \eqref{rottaca} is equal to $\log k$  when $\max_{x\in \mathcal X} p(x| Y=y ) = \frac{1}{k}$
 for $k\in \{1,2,3,\ldots\}$, and it is also monotonically decreasing in $\max_{x\in \mathcal X} p(x| Y=y )$. 
 Hence, if $\max_{x\in \mathcal X} p(x| Y=y ) \in  (\frac{1}{k+1} , \frac{1}{k}]$, then the lower bound 
 on $H_\alpha(X | Y=y)$, in the right-hand side of \eqref{rottaca}, lies in the interval $[\log k, \log(k + 1))$, from which it follows that 
 \eqref{rottaca}  is equivalent to state the following theorem \cite{Sason_Verdu_MHyp_2018}\footnote{ We note that Theorem~\ref{th:Sason_Verdu_MHyp_2018} is similar to \cite[Theorem~12]{Sason_Verdu_MHyp_2018} except that the latter is given in terms of the Arimoto-R\'{e}nyi conditional entropy, $H_{\alpha}(X|Y)$, whereas the former is given in terms of the R\'{e}nyi entropy of the random variable  $X$ conditioned to the event ${Y=y}$, i.e $H_{\alpha}(X|Y=y)$.}:
 
\begin{theo}[\hspace{-0.1mm}\cite{Sason_Verdu_MHyp_2018}]
\label{th:Sason_Verdu_MHyp_2018}
Let $k\in\{1,2,3,\ldots\}$, and $\alpha \in (0,1) \cup (1, \infty)$. If $\log k \leq H_\alpha(X | Y=y) < \log (k+1)$, then
\begin{IEEEeqnarray}{lll}
& 1-\max_{x\in \mathcal X} p(x| y )  \nonumber \\ 
& \leq \!
\frac{\exp{ \{ \frac{1-\alpha}{\alpha}  H_\alpha(X | Y=y)  \} } - 
k^{\frac{1}{\alpha}+1} +(k-1)(k+1)^{\frac{1}{\alpha}} }
{ k(k+1)^{\frac{1}{\alpha}} - k^{\frac{1}{\alpha}}(k+1)}\!. \quad\,\,
\label{rottaca1}
\end{IEEEeqnarray}
Furthermore, the upper bound on $1-\max_{x\in \mathcal X} p(x| y ) $ as a function of 
$ H_\alpha(X | Y=y)$ is  asymptotically tight in the limit where $\alpha \rightarrow \infty$. 
\end{theo}

Next, for a given graph realization $\bf E = \Edg$, let $\Lset^{N} =\{\Lbl_{(1)}, \Lbl_{(2)}, \ldots, \Lbl_{(|\Lset|^{N})}\}$ denote the  set of all possible label estimates ordered according to their posterior probability (i.e.  normalized accuracy), as per Definition \ref{order}.
Note that conditioned on $\Edg$, the probability of correct classification of the jointly optimal MAP classifier and active learning policy, $P_c^{\mathrm{opt,ssp}}(\Edg)$, can be always 
lower bounded as
\begin{IEEEeqnarray}{lll}\label{Eq:LowAppendix}
	P_c^{\mathrm{opt,ssp}}(\Edg) \geq  \sum_{i=1}^{\gamma}  f(\Lbl_{(i)}|\Edg)
\end{IEEEeqnarray}
with a properly chosen constant $\gamma$.
A valid value for $\gamma$ is  $M+1$.  To see this, let  $P_c^{\mathrm{\pi,ssp}}(\Edg)$
be the the probability of correct classification, conditioned on $\Edg$, 
of  the semi-supervised MAP classifier with a \textit{suboptimal}  batch query selection policy $\pi$.
Note that similar to $P_c^{\mathrm{opt,ssp}}(\Edg)$, given by  the sum of $|\Lset|^M$  proper posteriors, also 
$P_c^{\mathrm{\pi,ssp}}(\Edg)$ can be written in terms of the sum of $|\Lset|^M$ posteriors.
Next, note that we can always design 
the  batch  policy $\pi$ such 
its $M$ queries  are chosen to include in the sum the first $M+1$ largest posteriors. Therefore,  
 a lower bound on $P_c^{\mathrm{\pi,ssp}}(\Edg)$, and consequently on  $P_c^{\mathrm{opt,ssp}}(\Edg)$, 
 is given by \eqref{Eq:LowAppendix} with $\gamma=M+1$ neglecting the remaining $|\Lset|^M-M-1$ posteriors. 
For posterior probabilities that are {\em label permutation invariant}, (see Definition~\ref{Def:PermInv} for details), a tighter lower bound can be obtained from \eqref{Eq:LowAppendix} by choosing $\gamma$ as
$ |\Lset|!(M-|\Lset|+2)$ if $ M \geq |\Lset| -1$ and $\frac{ |\Lset|!}{( |\Lset|-M)!} $ otherwise. Using Theorem~\ref{th:Sason_Verdu_MHyp_2018}, the rest of the proof is to obtain a lower bound on the ordered posteriors $f(\Lbl_{(i)}|\Edg)$ in terms of R\'{e}nyi~Entropy of properly defined random variables.

As already done in Appendix \ref{App:Renyi_upperbound},  let us define the following conditional variable:  $X_{|\Edg}$ is a mapping  that goes from $\Lset^{N} =\{\Lbl_{(1)}, \Lbl_{(2)}, \ldots, \Lbl_{(|\Lset|^{N})}\}$ to   $\mathcal X =\{1, \dots, |\Lset|^{N}\}$  defined as  $X_{|\Edg}(\Lbl_{(i)})= i$. Therefore,  $P(X_{|\Edg}=i)= f(\Lbl_{(i)}|\Edg)$ follows. Applying \eqref{rottaca1}  in Theorem \ref{th:Sason_Verdu_MHyp_2018} to $X_{|\Edg}$ and recalling that 
$\max_{i \in \mathcal X} P(X_{|\Edg}=i)= f(\Lbl_{(1)}|\Edg)$ it follows that
\begin{eqnarray}
 f(\Lbl_{(1)}|\Edg) \geq f^{\mathrm{LB}}(\Lbl_{(1)}|\Edg)  \label{Eq:Flb1} 
\end{eqnarray}
with
\begin{IEEEeqnarray}{lll}
	f^{\mathrm{LB}}(\Lbl_{(1)}|\Edg)  = \nonumber \\
	1-  \!\!    \!\!\!\! \inf_{(k,\alpha) \in \mathcal D}  \!\!\!\!  \frac{\exp{ \{ \frac{1-\alpha}{\alpha}  H_\alpha(X_{|\Edg})  \} }\! - \!
		k^{\frac{1}{\alpha}+1} \!+\! (k \!-\! 1)(k \!+\! 1)^{\frac{1}{\alpha}} }
	{ k(k+1)^{\frac{1}{\alpha}} - k^{\frac{1}{\alpha}}(k+1)
	}, \nonumber
\end{IEEEeqnarray}
where $\mathcal D=\{(k, \alpha): \, k   \in \mathbb{S}, \, \alpha \in (0,1)\cup (1,\infty) \}$. Let us now define a new conditional random variable $X^{(1)}_{|\Edg}$ defined by the following conditional probability mass function 
 \begin{eqnarray}
P(X^{(1)}_{|\Edg}=i)=\begin{cases}
\frac{f(\Lbl_{(1)}|\Edg) -f^{\mathrm{LB}}(\Lbl_{(1)}|\Edg)}{1 - f^{\mathrm{LB}}(\Lbl_{(1)}|\Edg)} & \mbox{for}  \, i =1 \\
\frac{f(\Lbl_{(i)}|\Edg)}{1 - f^{\mathrm{LB}}(\Lbl_{(1)}|\Edg)} & \mbox{for}  \, i \in \{2, \ldots, |\Lset|^{N} \} 
\end{cases}
\nonumber 
\end{eqnarray}
Note that the 
\begin{IEEEeqnarray}{lll}
\max_{i \in \mathcal X} P(X^{(1)}_{|\Edg}=i) \nonumber \\
=\frac{ \max\left\{ f(\Lbl_{(2)}|\Edg), f(\Lbl_{(1)}|\Edg) - f^{\mathrm{LB}}(\Lbl_{(1)}|\Edg) \right\} }{1 - f^{\mathrm{LB}}(\Lbl_{(1)}|\Edg)}. \nonumber
\end{IEEEeqnarray}
Hence,  applying  Theorem \ref{th:Sason_Verdu_MHyp_2018} to  $X^{(1)}_{|\Edg}$, we have 
\begin{IEEEeqnarray}{lll}
\max\left\{ f(\Lbl_{(2)}|\Edg), f(\Lbl_{(1)}|\Edg) - f^{\mathrm{LB}}(\Lbl_{(1)}|\Edg) \right\}
 \nonumber \\
  \geq  (1 - f^{\mathrm{LB}}(\Lbl_{(1)}|\Edg))f^{\mathrm{LB}}(\Lbl_{(2)}|\Edg) \label{Eq:Flb2} 
\end{IEEEeqnarray}
with
\begin{IEEEeqnarray}{lll}
	f^{\mathrm{LB}}(\Lbl_{(2)}|\Edg)  =
	\nonumber \\
	1-  \!\!\!\!   \inf_{(k,\alpha) \in \mathcal D}  \!\!\!\!    \frac{\exp{ \{ \frac{1-\alpha}{\alpha}  H_\alpha(X^{(1)}_{|\Edg})  \} } \!-\! 
		k^{\frac{1}{\alpha}+1} \!+\! (k \!-\!1)(k \!+\! 1)^{\frac{1}{\alpha}} }
	{ k(k+1)^{\frac{1}{\alpha}} - k^{\frac{1}{\alpha}}(k+1)
	}
	\nonumber 
\end{IEEEeqnarray}
with $H_\alpha(X^{(1)}_{|\Edg})$ denoting the  R\'{e}nyi Entropy of $X^{(1)}_{|\Edg}$. 
Hence, it is immediate that  a lower bound on the sum of the first two posteriors can be obtained as: 
\begin{IEEEeqnarray}{lll}
	f^{\mathrm{LB}}(\Lbl_{(1)}|\Edg) + (1 - f^{\mathrm{LB}}(\Lbl_{(1)}|\Edg))f^{\mathrm{LB}}(\Lbl_{(2)}|\Edg) \nonumber \\
	 { \overset{(a)}{\leq}}  f^{\mathrm{LB}}(\Lbl_{(1)}|\Edg)  +  \max\left\{ f(\Lbl_{(2)}|\Edg),  f(\Lbl_{(1)}|\Edg) - f^{\mathrm{LB}}(\Lbl_{(1)}|\Edg) \right\} 
	\nonumber \\ 
	= \begin{cases}
		f(\Lbl_{(1)}|\Edg), \quad \text{if}\,\, f(\Lbl_{(2)}|\Edg) \leq  f(\Lbl_{(1)}|\Edg) - f^{\mathrm{LB}}(\Lbl_{(1)}|\Edg) \\
		f^{\mathrm{LB}}(\Lbl_{(1)}|\Edg)  + f(\Lbl_{(2)}|\Edg), \quad \text{otherwise} 
		\end{cases}
	\nonumber \\ 
	{ \overset{(b)}{\leq}} 	f(\Lbl_{(1)}|\Edg) + f(\Lbl_{(2)}|\Edg),
\end{IEEEeqnarray}
 where inequality $(a)$ follows from \eqref{Eq:Flb2} and  inequality $(b)$ follows from $f(\Lbl_{(2)}|\Edg)\geq 0$ if $f(\Lbl_{(2)}|\Edg) \leq  f(\Lbl_{(1)}|\Edg) - f^{\mathrm{LB}}(\Lbl_{(1)}|\Edg)$ holds and from \eqref{Eq:Flb1}, otherwise. 

In order to obtain a lower bound on the first $\gamma$ posteriors, we will adopt in an iterative way what we have applied
to  $X^{(1)}_{|\Edg}$.
Specifically,  let  $X^{(n)}_{|\Edg}$ be a random variable defined in the following iterative way:
$X^{(0)}_{|\Edg}=X_{|\Edg}$ and  for all $n=1, \ldots, \gamma$,
\begin{eqnarray}
P(X^{(n)}_{|\Edg}=i) =\frac { \theta_i^{(n)}}{ 1 - \sum_{j=1}^{n} \widetilde{\mathcal{B}}^{(j)} }
\label{landed}
\end{eqnarray}
where, for all $i  \in \{1, \ldots,  |\Lset|^{N} \}$, $\theta_i^{(0)}= f(\Lbl_{(i)}|\Edg)$ while 
$\theta_i^{(n)}$ is defined as:
 \begin{eqnarray}
\theta_i^{(n)}  \!=\! 
\left \{\begin{array}{lll}
\theta_i^{(n-1)} -   \widetilde{\mathcal{B}}^{(n)} 
\, \,  & \text{for}  \,  \, \, i = i^* \\
\displaystyle \theta_i^{(n-1)}   &  \text{for}  \,  \, \, i\neq i^*   \in \{1, \ldots,  |\Lset|^{N} \}
\end{array}
\right.
\label{landed2}
\end{eqnarray}
where $$
i^*= \arg\max_{i  \in \{1, \ldots,  |\Lset|^{N} \} } \left\{\theta_i^{(n-1)} \right\}
$$
and
\begin{IEEEeqnarray}{lll}
\BLUE	\widetilde{\mathcal{B}}^{(n)}= f^{\mathrm{LB}}(\Lbl_{(n)}|\Edg)(1-\sum_{j=1}^{n-1} \widetilde{\mathcal{B}}^{(j)}),
\label{landed3}
\end{IEEEeqnarray}
 with 
 $\sum_{j=1}^{0} \widetilde{\mathcal{B}}^{(j)}=0$ by convention,
\begin{eqnarray}\label{Eq:FlB}
 \!\!\!\! \!\!\!\!  \!\!\!\!  && f^{\mathrm{LB}}(\Lbl_{(n)}|\Edg)  =   \\ 
\!\!\!\!  \!\!\!\!  \!\!\!\!  &&  1-  \!\!\!\!   \inf_{(k,\alpha) \in \mathcal D}  \!\!\!\!    \frac{\exp{ \{ \frac{1-\alpha}{\alpha}  H_\alpha(X_{|\Edg}^{{\BLUE (n-1)}})  \} } - 
k^{\frac{1}{\alpha}+1} +(k-1)(k+1)^{\frac{1}{\alpha}} }
{ k(k+1)^{\frac{1}{\alpha}} - k^{\frac{1}{\alpha}}(k+1)
} \nonumber 
\end{eqnarray}
and  $H_\alpha(X^{(n)}_{|\Edg})$ denoting  R\'{e}nyi Entropy of $X^{(n)}_{|\Edg}$.

 Then,  
 a lower bound on the sum of the $i$-th largest posteriors, $f(\Lbl_{(i)}|\Edg)$  is given by:
\begin{IEEEeqnarray}{lll}\label{Eq:Post_Beta}
 \sum_{i=1}^{n}  f(\Lbl_{(i)}|\Edg)\geq \sum_{i=1}^{n} \BLUE \widetilde{\mathcal{B}}^{(i)},
 \end{IEEEeqnarray}
from which, using 
{\BLUE \eqref{Eq:LowAppendix}}, Theorem~\ref{Renyi_lowerbound}
follows immediately after proving that $\forall \alpha \in (1, +\infty)$
 \begin{IEEEeqnarray}{lll}\label{Eq:bound_Renyi}
\exp{ \left \{ \frac{1-\alpha}{\alpha}  H_\alpha(X^{(i)}_{|\Edg}) \right \} } \!\leq \!
\exp{ \left \{ \frac{1-\alpha}{\alpha}  H^{(i)}_\alpha({\bf L}|{\bf E}=\Edg) \right \}} \quad\,\,\,\,
\end{IEEEeqnarray}
{\BLUE with $H^{(i)}_\alpha({\bf L}|{\bf E}=\Edg) $ defined in \eqref{iteration}. Note that for $i=1$, we have $H^{(1)}_\alpha({\bf L}|{\bf E}=\Edg)=H_\alpha({\bf L}|{\bf E}=\Edg) $.} 

In order to prove \eqref{Eq:bound_Renyi},  we prove that $H_\alpha(X^{(i)}_{|\Edg})\geq H^{(i)}_\alpha({\bf L}|{\bf E}=\Edg) $ holds for $\forall \alpha \in (1, +\infty)$. To this end, it is enough to observe that given a probability mass vector $\bf p=[p_1, \ldots, p_{|\mathcal X|}]$ with $p_1\geq p_2 \geq\cdots \geq p_{|\mathcal X|}$
and given a probability mass vector $\bf p'$ such that
 \begin{IEEEeqnarray}{lll}
	p'_i = \begin{cases}
		\frac{p_i-\kappa_i}{1-\sum_{i=1}^{j}\kappa_j}\quad \text{for}\,\,i\leq j\\
		\frac{p_i}{1-\sum_{i=1}^{j}\kappa_j}\quad \text{for}\,\,i> j,
		\end{cases}
\end{IEEEeqnarray}
 the following bound holds for  $\alpha \in (1,+\infty)$
 \begin{IEEEeqnarray}{lll}\label{Eq:RenEntUpp}
H_\alpha({\bf p'})  =  \frac{1}{1-\alpha} \log \sum_{i\in \mathcal X}  {p'_i}^\alpha \nonumber
\\ 
 =  \frac{1}{1-\alpha} \log \bigg(\frac{1}{(1-\kappa)^{\alpha}}\sum_{i\in \mathcal X}  {p_i}^\alpha \nonumber \\
\qquad\qquad\qquad  - \frac{\sum_{i=1}^j p_i^{\alpha}}{(1-\kappa)^{\alpha}} + \frac{\sum_{i=1}^j(p_i-\kappa_i)^{\alpha}}{(1-\kappa)^{\alpha}} \bigg) \nonumber
\\ 
 =  \frac{1}{1-\alpha} \bigg[-\alpha \log(1-\kappa) \nonumber \\
 \,\,\,\, +\log\bigg(\exp\Big\{\log\sum_{i\in \mathcal X}  {p_i}^\alpha \Big\} - \sum_{i=1}^j [p_i^{\alpha} - (p_i-\kappa_i)^{\alpha}] \bigg) \bigg] \nonumber
\\ 
 =  \frac{1}{1-\alpha} \bigg[-\alpha \log(1-\kappa) \nonumber \\
 \,\,\,\, +\log\bigg(\exp\Big\{(1-\alpha)H_\alpha({\bf p}) \Big\} - \sum_{i=1}^j [p_i^{\alpha} - (p_i-\kappa_i)^{\alpha}] \bigg) \bigg] \nonumber
\\ 
 \overset{(a)}{\geq}
 \frac{1}{1-\alpha} \bigg[-\alpha \log(1-\kappa) \nonumber \\
 \,\,\,\, +\log\bigg(\exp\Big\{(1-\alpha)H_\alpha({\bf p}) \Big\} - \sum_{i=1}^j \kappa_i^{\alpha} \bigg) \bigg],
\end{IEEEeqnarray}
where $\kappa=\sum_{i=1}^{j}\kappa_i$ and inequality $(a)$ follows from $ p_i^{\alpha} - (p_i-\kappa_i)^{\alpha}\geq \kappa_i^{\alpha}$. 



Using  \eqref{Eq:bound_Renyi} and recalling ${\mathcal{B}}^{(i)}$ defined in Theorem~\ref{Renyi_lowerbound}, we have that ${\mathcal{B}}^{(1)}= \widetilde{\mathcal{B}}^{(1)} $ and
${\mathcal{B}}^{(2)}\leq \widetilde{\mathcal{B}}^{(2)}$. Furthermore,  using in  \eqref{landed}-\eqref{landed3}, for all $i = 1, \ldots, \gamma$, 
${\mathcal{B}}^{(i)}$ rather then  $\widetilde{\mathcal{B}}^{(i)}$, it is immediate to prove, following  arguments similar to the ones used  for 
$\widetilde{\mathcal{B}}^{(i)}$, that
\begin{IEEEeqnarray}{lll} 
	P_c^{\mathrm{opt,ssp}}(\Edg) \geq  \sum_{i=1}^{\gamma}  f(\Lbl_{(i)}|\Edg) 
	\geq  \sum_{{ i=1}}^{\gamma }  [\mathcal{B}^{(i)}]^+, \quad
\end{IEEEeqnarray}
which completes the proof.




%

\section{Proof of Theorem~\ref{Theo:LowerMGaingeneral}}\label{App:TheoLowerMGaingeneral}

In the following, we prove the   upper and lower bounds provided in 
Theorem~\ref{Theo:LowerMGaingeneral} 
 on the relative gain of the jointly optimal MAP classifier and active learning  policy versus unsupervised MAP classifier.  For a given observable realization
$\Edg$,  let $\bf L^{\mathrm{opt}}_{\Nlb}$ (and $\Lbl^{\mathrm{opt}}_{\Nlb}$)
the  random vector (and its realization) resulting from the optimal query policy. 

For a given observable realization $\Edg$, following the derivation provided in Appendix \ref{App:Renyi_upperbound} (i.e. \eqref{triste011}), it follows that   $P_c^{\mathrm{opt,ssp}}(\Edg)$  is given by:
\begin{IEEEeqnarray}{lll}
P_c^{\mathrm{opt,ssp}}(\Edg)\, &=\mathbb{E}\Big[ P_c^{\mathrm{opt,ssp}}({\bf E},{\bf L}^{\mathrm{opt}}_{\Nlb})  | {\bf E}= \Edg \Big]  \nonumber \\
&= \sum_{\Lbl_{\Nlb} \in\mathcal P }   {\max}_{\Lbl_{\Nul}}\,\,f(\Lbl_{\Nul},\Lbl_{\Nlb}|\Edg). 
\label{triste011}
\end{IEEEeqnarray}
Therefore, $P_c^{\mathrm{opt,ssp}}(\Edg)$ is in general the sum of $|\Lset|^M$ posterior probabilities. 
From \eqref{triste011}, it follows  immediately that  an upper bound for $P_c^{\mathrm{opt,ssp}}(\Edg)$ is the sum of the $|\Lset|^M$ largest posterior probabilities i.e.
\begin{eqnarray}
P_c^{\mathrm{opt,ssp}}(\Edg) \leq  \sum_{i=1}^{{|\mathcal L|}^M}  f(\Lbl_{(i)}|\Edg)
\label{triste}
\end{eqnarray}
from which after normalizing   \eqref{triste} by $P_c^{\mathrm{usp}}(\Edg)=f(\Lbl_{\mathcal N}^*|\Edg) $, the upper bound given in \eqref{Eq:BoundSupUnsupEdg1} is obtained. 

For the lower bound, we consider the following suboptimal batch query selection. Let $\Gamma$ be the largest  index of the ordered label estimates 
such that there exist
$M$ items whose label configurations corresponding to the first $\Gamma$ posterior probabilities are distinct. The considered batch query selection policy chooses the aforementioned $M$ data items which implies that the $|\Lset|^M$ terms in the summation in \eqref{triste011} include at least the $\Gamma$ largest posteriors. This immediately leads to the following lower bound 
\begin{eqnarray}
P_c^{\mathrm{opt,ssp}}(\Edg) \geq  \sum_{i=1}^{\Gamma}  f(\Lbl_{(i)}|\Edg)
\label{tristu}
\end{eqnarray}
from which after normalizing   \eqref{tristu} by $P_c^{\mathrm{usp}}(\Edg)=f(\Lbl_{\mathcal N}^*|\Edg)$, the lower bound given in \eqref{Eq:BoundSupUnsupEdg1} directly follows.  This concludes the proof.






\bibliographystyle{IEEEtran}
\bibliography{References,References_project,references,QualRef,EE,references_d2d_2,NSF_CIF_Privacy,ref,refabbas,references,references2}

%
%
%
%
%
%
%
%

\end{document}